\documentclass{IEEEtran}
\usepackage{cite}
\usepackage{amsmath,amssymb,amsfonts,bm}
\usepackage{algorithmic}
\usepackage{graphicx}
\usepackage{booktabs}
\usepackage{textcomp}
\usepackage{algorithm}
\usepackage{makecell}
\usepackage{graphicx}
\usepackage{subcaption}
\usepackage{csquotes}
\usepackage{multirow}
\usepackage{subcaption}
\def\BibTeX{{\rm B\kern-.05em{\sc i\kern-.025em b}\kern-.08em
    T\kern-.1667em\lower.7ex\hbox{E}\kern-.125emX}}

\begin{document}
\title{WEKP-PLP: A Wireless Environment Knowledge Pool-Enhanced Path Loss Prediction Framework \\for Shore-to-Ship Communication}
\author{Jikun Du,
Lei Tian,~\IEEEmembership{Member, IEEE},
Jianhua Zhang,~\IEEEmembership{Fellow, IEEE}, Pan Tang,~\IEEEmembership{Member, IEEE}, 
Zihang Ding, Bin Ao, Zhen Zhang~\IEEEmembership{Member, IEEE}, Jialin Wang
and Yuanzhi He
\thanks{This work was supported by National Natural Science Foundation of China (62525101, 62571059 and 62571053), Natural Science Foundation of Beijing-Xiaomi Innovation Joint Foundation (L243002), Guangdong Major Project of Basic and Applied Basic Research under Grant
2023B0303000001, and Beijing University of Posts and Telecommunications-China Mobile Communications Group Co.,Ltd. Joint Institute. (Corresponding author: Lei Tian.)}
\thanks{J. Du, L. Tian, J. Zhang, P. Tang, Z. Ding, and B. Ao are with the State Key Laboratory of Networking and Switching Technology, Beijing University of Posts and Telecommunications, Beijing 100876, China (e-mail: \{dujikun0904, tianlbupt, jhzhang, tangpan27, dingzihang, binao\}@bupt.edu.cn).}
\thanks{Z. Zhang is with the Inner Mongolia Key Laboratory of Intelligent Communication and Sensing and Signal Processing, Inner Mongolia University, Hohhot 010021, China (e-mail: zhenzhang@imu.edu.cn).}
\thanks{J. Wang is with the China Information Technology Designing \& Consulting Institute Co., Ltd., Beijing 100048, China (e-mail: wangjl733@chinaunicom.cn).}
\thanks{Yuanzhi He is with the 32008 Troops of PLA, Beijing 100141, China (e-mail: he\_yuanzhi@126.com).}}

\maketitle

\begin{abstract}
Accurate shore-to-ship path loss prediction is essential for maritime mobile communication systems, but it remains challenging because dynamic sea surface conditions can alter reflection paths and multipath effects, introducing uncertainty into the observed path loss.
This paper proposes a wireless environment knowledge
pool-enhanced path loss prediction (WEKP-PLP) method for point prediction and interval characterization of shore-to-ship path loss.
The proposed method first constructs a wireless environment knowledge pool (WEKP) from ray tracing (RT) simulations under different wind speeds, temperatures, salinities, frequencies, antenna heights, and propagation distances.
The WEKP learns the mapping from environmental and system parameters to path loss residual quantiles and provides a prior that contains both the median prediction and the associated prediction interval.
To adapt this prior to real scenarios, a small number of measurement samples are used to learn the residual between the WEKP median prediction and the measured path loss through Gaussian process regression (GPR).
The experimental results show that the WEKP-PLP provides accurate point prediction and compact intervals with reliable coverage.
The proposed method achieves an MAE of 1.03 dB and an RMSE of 1.54 dB.
The ablation results further confirm that both the WEKP prior and the residual correction using a small number of measurement samples are necessary for accurate and reliable shore-to-ship path loss prediction.
\end{abstract}


\section{Introduction}
The rapid growth of maritime services, such as maritime Internet of Things and marine transportation, is driving increasing demand for mobile communication networks with wider coverage, higher data rates, and improved reliability \cite{jianhua_2023}. 
Nearshore regions concentrate a large amount of maritime economic activity and communication demand, where scenarios such as smart ports and environmental monitoring require stable and reliable wireless connectivity \cite{Bose_2025, zhang_2024}. 
Compared with satellite communication, shore-to-ship communication relies on coastal mobile communication facilities and can provide low-cost and high-rate broadband services for nearshore users \cite{Xylouris_2024} . 
Therefore, accurately characterizing the shore-to-ship wireless propagation environment and constructing path loss models suitable for complex shore-to-ship scenarios are fundamental to the design and performance evaluation of shore-to-ship communication systems.

However, due to the unique characteristics of maritime propagation environment, shore-to-ship path loss modeling still faces several challenges \cite{wang_2018}. First, measurement data plays an important role in constructing reliable shore-to-ship channel models. They directly reflect the path loss characteristics in real maritime propagation environment and provide an important basis for model validation. However, existing maritime measurement data usually covers only limited environmental conditions, and large-scale datasets collected under diverse sea state conditions remain scarce \cite{tian_2026}.
In addition, the dynamic sea surface affects multipath propagation. In shore-to-ship channels, the direct path and sea surface reflected components are usually the dominant propagation components. Due to sea surface fluctuations, the positions of the effective reflection points may vary over time, which further changes the propagation lengths of the reflected components. As these reflected propagation lengths change, the phase relationships between the direct path and the reflected components also vary, thereby affecting their coherent superposition at the receiver. This phase variation induced by sea surface dynamics can lead to noticeable path loss fluctuations. Therefore, shore-to-ship path loss should be characterized with prediction intervals to explicitly represent the uncertainty induced by dynamic sea surface conditions. Such interval information can provide additional reliability information for link budget design and coverage evaluation in maritime communication systems.

Existing shore-to-ship channel modeling approaches can be broadly categorized into statistical, deterministic, and artificial intelligence (AI)-based channel modeling methods \cite{wangh_2025}. Statistical channel models usually represent the channel through empirical functions or statistical distributions. The two-ray model is widely used for maritime path loss calculation by considering the direct path and the sea surface reflected path under idealized reflection conditions. To improve its accuracy, several studies have incorporated additional effects such as sea surface roughness \cite{mi_2025}, shadowing \cite{sun_2026}, Earth curvature \cite{yang_2019}, and evaporation duct \cite{wang_2025}. However, two-ray models usually assume fixed antenna heights, whereas sea surface fluctuations may change the effective antenna height, and neglecting this effect may lead to considerable modeling errors \cite{jiaqi_2025}.

Deterministic channel models, especially RT, can explicitly describe the interactions between radio waves and the propagation environment. Specifically, \cite{ding_2019} proposed a stochastic RT method for maritime channels, which simplified sea wave effects into positional changes of transceivers and reflection points. Furthermore, \cite{chen_2022, wang_2021} used the Pierson-Moskowitz (PM) wave spectrum and RT method for offshore electromagnetic propagation analysis, which achieved an accurate characterization of complex maritime environments. Deterministic channel models can explicitly characterize the main propagation mechanisms in shore-to-ship channels. However, the number of discretized sea surface grids increases rapidly with simulation distance, leading to high computational cost and limiting their direct application to large-scale shore-to-ship path loss modeling.

AI-based channel modeling learns the mapping from input features to channel characteristic parameters, thereby improving adaptability to varying scenarios \cite{yu_2025}.
For example, a two-stage prediction framework was proposed by combining a two-ray model with a feedforward neural network, where the neural network learns the residuals between the two-ray model and the measured path loss \cite{noh_2024}.
Additionally, \cite{zhangh_2024} introduced an unsupervised clustering approach to build a segmented path loss model combining the CI model with Rician fading. 
\cite{peng_2025} addressed the prediction of 5G reference signal received power based on a self-developed maritime 5G terminal. 
These studies suggest that AI methods can be used not only for channel parameter prediction, but also for model parameter optimization. However, AI-based models usually require a large amount of measurement data for training, which is difficult to obtain in maritime environments because measurement campaigns are less accessible than those in terrestrial scenarios.

Despite recent progress, two issues remain insufficiently addressed for shore-to-ship path loss prediction. First, available maritime measurement data are usually limited in spatial and environmental coverage, making it difficult to support reliable parameter fitting, model validation, and AI-based model training. Second, most existing methods mainly focus on deterministic point prediction, which limits their ability to characterize the uncertainty and fluctuation range of path loss under dynamic sea surface conditions.

To address these challenges, this paper proposes WEKP-PLP, a wireless environment knowledge pool-enhanced path loss prediction framework for shore-to-ship communication. The proposed framework first constructs a WEKP from RT simulation data obtained under different environmental and system conditions, including wind speed, seawater temperature, salinity, frequency, antenna height, and propagation distance. Different from the WEKP studies in \cite{jialin_2024, jialin_2025}, which mainly focus on quantifying the contributions of propagation mechanisms to channel characteristics, this work uses the WEKP to learn the relationship between maritime environmental and system conditions and the quantiles of the residual between the RT-simulated path loss and the FSPL baseline. By adding the FSPL baseline, the learned residual quantiles are transformed into path loss quantiles, which provide both point prediction and uncertainty characterization through prediction intervals.
To adapt the WEKP-based path loss quantile prior to real shore-to-ship environment, a small number of measurement samples are further used for measurement-based residual correction. Specifically, the correction model learns the discrepancy between the measured path loss and the WEKP-based median prediction, thereby compensating for propagation effects that are not fully represented in the RT simulation. Therefore, WEKP-PLP combines the WEKP-based path loss quantile prior with measurement-based residual correction using a small number of measurement samples, thereby providing final point prediction and interval characterization.
The main contributions of this paper are summarized as follows.

\begin{itemize}
\item A shore-to-ship path loss prediction framework with interval characterization is formulated. Unlike conventional prediction methods, the proposed WEKP-PLP framework provides both a point estimate and a prediction interval, enabling the uncertainty of path loss under dynamic maritime conditions to be explicitly characterized.

\item A WEKP is proposed to establish the mapping from environmental and system conditions to the quantiles of the residual between the RT-simulated path loss and the FSPL baseline. In this work, the WEKP is implemented using Categorical Boosting (CatBoost) quantile regression. This provides a prior from RT simulation data and reduces the requirement for large-scale measurement data.

\item A measurement-based residual correction method using GPR is developed for real shore-to-ship environment. Since the RT simulation mainly considers the direct path and the sea surface reflected components, practical measurements may include additional unmodeled propagation effects. A small number of measurement samples are used to train the GPR model, which learns the discrepancy between the measured path loss and the WEKP-based median prediction. This correction improves point prediction accuracy while maintaining reliable interval coverage.
\end{itemize}

The rest of this paper is organized as follows. Section~\ref{sec:2} describes the shore-to-ship propagation scenario and formulates the point prediction and interval characterization problem. Section~\ref{sec:3} presents the construction of WEKP and the corresponding quantile prediction formulation. Section~\ref{sec:4} introduces the  correction method for adapting the WEKP to real-world shore-to-ship prediction. Section~\ref{sec:5} provides the experimental results and discussion. Finally, Section~\ref{sec:6} concludes this paper.

\section{System Model and Problem Formulation} \label{sec:2}
This section formulates the shore-to-ship path loss  prediction problem considered in this paper. 
In shore-to-ship mobile communication, the received signal is mainly affected by the direct path and the sea surface reflected components, while sea surface variations may change their phase relationship and cause path loss fluctuations. 
Therefore, a single deterministic prediction is insufficient to describe such fluctuation behavior.
Given the environmental and system parameters, this work aims to predict the lower, median, and upper quantiles of path loss. 
The median quantile is used as the predicted path loss, and the lower and upper quantiles define the prediction interval.

\subsection{Shore-to-Ship Propagation Scenario}
\label{sub:2.1}
This paper considers the shore-to-ship propagation scenario, where the transmitter is deployed on the coast and the receiver is located on a vessel over the sea surface. 
Compared with dense terrestrial environments, the shore-to-ship scenarios usually have a more open propagation space. 
In this scenario, the received signal is mainly affected by the direct path and the sea surface reflected components.
In addition, the rough sea surface may introduce sea surface scattering.
Meanwhile, it can also be influenced by coastal buildings, harbor facilities, and other scatterers near the shoreline.
These objects may introduce additional reflected, diffracted, and scattered paths. 

\begin{figure}[t]
\centering
\begin{subfigure}{0.9\columnwidth}
    \centering
    \includegraphics[width=\linewidth]{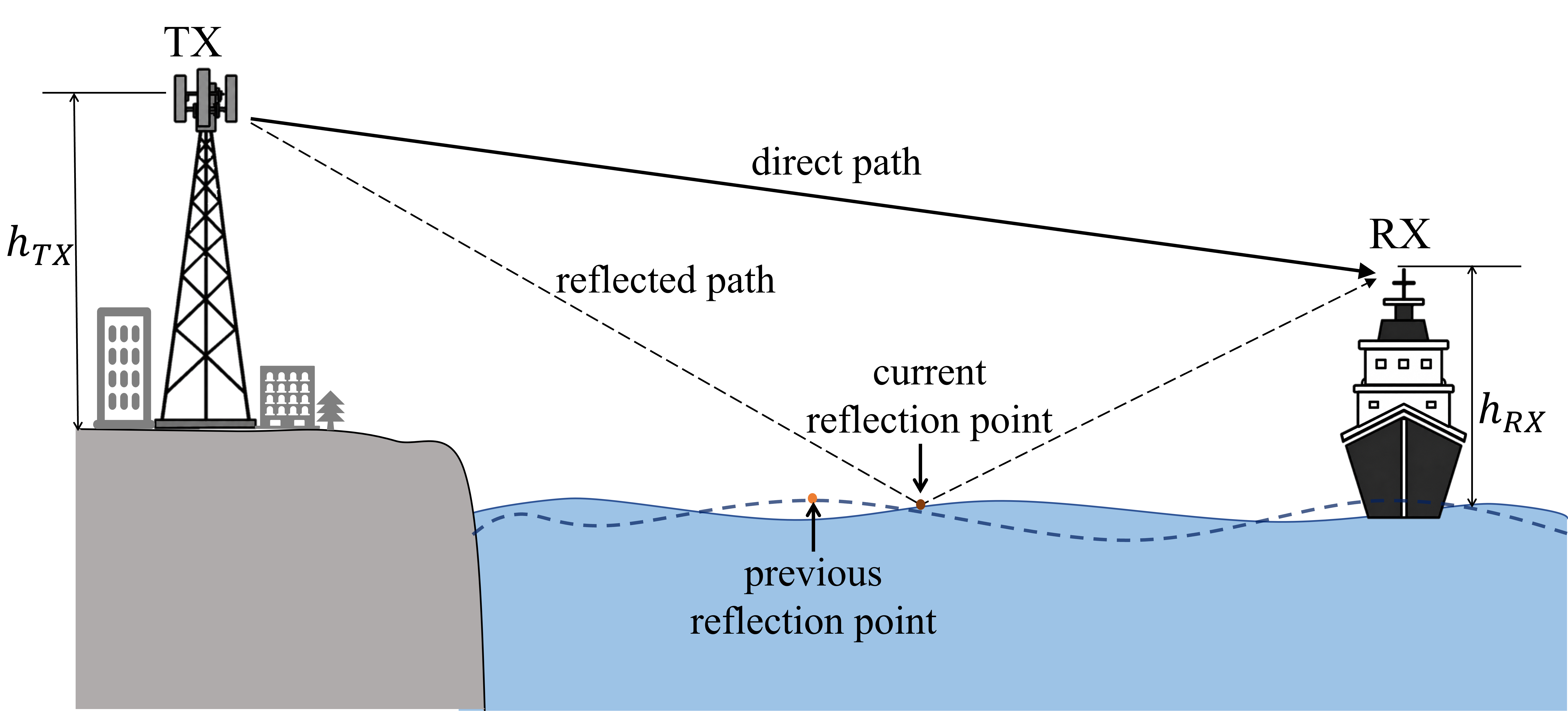}
    \caption{}
    \label{fig:sea_a}
\end{subfigure}
\begin{subfigure}{0.9\columnwidth}
    \centering
    \includegraphics[width=\linewidth]{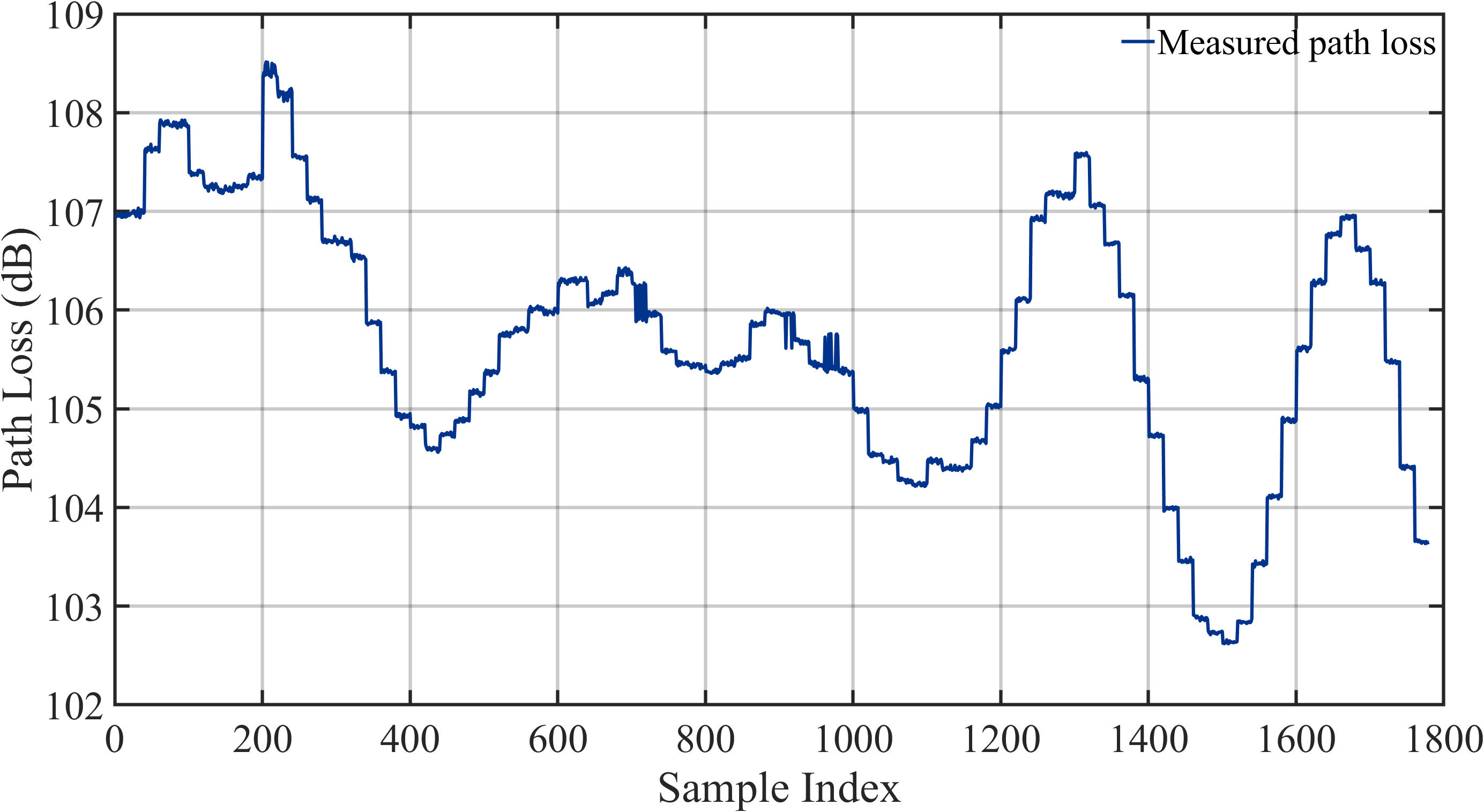}
    \caption{}
    \label{fig:sea_b}
\end{subfigure}
\caption{Illustration of path loss fluctuations in shore-to-ship propagation. (a) Variation of the sea surface reflection point caused by dynamic sea surface conditions. (b) Short-term path loss variation.}
\label{fig:sea_channel}
\end{figure}

In shore-to-ship scenarios, environmental conditions such as wind speed can affect the sea surface state, thereby altering the local height and slope of the sea surface.
Since sea surface reflected components often act as major propagation components in shore-to-ship scenarios, these changes may shift the effective reflection points and alter the reflected path lengths.
Consequently, the phase differences between the direct path and the sea surface reflected components become time varying, which changes their coherent superposition and leads to path loss fluctuations.
Fig.~\ref{fig:sea_channel} provides an illustration of this phenomenon. 
Fig.~\ref{fig:sea_channel}(\subref{fig:sea_a}) shows the direct path and the sea surface reflected path, together with the shift of the reflection point caused by the dynamic sea surface. 
Fig.~\ref{fig:sea_channel}(\subref{fig:sea_b}) gives an example of measured path loss variation within one minute, during which the vessel remains nearly stationary and the propagation distance changes by less than 1 m. The observed short-term fluctuations show that shore-to-ship path loss may vary around its point prediction value, suggesting the need to characterize shore-to-ship path loss using prediction intervals.

\subsection{Path Loss Prediction Formulation}\label{sub:2.2} 

The above analysis indicates that shore-to-ship path loss can exhibit the fluctuations due to the time-varying interaction between the direct path and the sea surface reflected components. A conventional prediction model provides a single representative path loss value, which is useful for estimating the propagation trend but is insufficient to describe the possible variation range around this value. To address this limitation, this paper retains point prediction as the primary objective and further introduces lower and upper quantiles to characterize the uncertainty of path loss. 

Let $\mathbf{x}$ denote the input feature vector, which is defined as
\begin{equation}
\mathbf{x} = [d, f, v, T, S],
\label{eq:input_vector}
\end{equation}
where $d$ denotes the propagation distance, $f$ denotes the carrier frequency, $v$ denotes the wind speed, $T$ denotes the seawater temperature, and $S$ denotes the seawater salinity. 
These variables describe system settings and maritime environmental conditions considered in this paper. 
Let $y$ denote the corresponding path loss. 
Instead of learning a deterministic mapping from $\mathbf{x}$ to a single value of $y$, the objective is to estimate the conditional quantiles of path loss for a given input feature vector.

For a quantile level $\tau \in (0,1)$, the conditional quantile of path loss is defined as
\begin{equation}
Q_{\tau}(\mathbf{x})
=
\inf \left\{
m: \mathbb{P}(y \leq m \mid \mathbf{x}) \geq \tau
\right\}
\label{eq:conditional_quantile}
\end{equation}
where $m$ is a candidate path loss value, and $\mathbb{P}(y \leq m \mid \mathbf{x})$ represents the conditional probability that the path loss is not larger than $m$ under the input feature vector $\mathbf{x}$.
For a prediction interval with a target coverage level of $1-\alpha$, where $\alpha$ denotes the significance level, the lower and upper bounds are obtained from the $\alpha/2$ and $1-\alpha/2$ quantiles, respectively.
Accordingly, the predicted path loss interval $\hat{I}_{1-\alpha}(\mathbf{x})$ is constructed as
\begin{equation}
\hat{I}_{1-\alpha}(\mathbf{x}) =
\left[
\hat{Q}_{\alpha/2}(\mathbf{x}),
\hat{Q}_{1-\alpha/2}(\mathbf{x})
\right].
\label{eq:prediction_interval}
\end{equation}
where $\hat{Q}_{\alpha/2}(\mathbf{x})$ and $\hat{Q}_{1-\alpha/2}(\mathbf{x})$ denote the predicted lower and upper quantiles, respectively.
The median quantile $\hat{Q}_{0.50}(\mathbf{x})$ is used as the predicted path loss.

The interval width $\Delta Q(\mathbf{x})$ is defined as
\begin{equation}
\Delta Q(\mathbf{x}) =
\hat{Q}_{1-\alpha/2}(\mathbf{x})
-
\hat{Q}_{\alpha/2}(\mathbf{x}).
\label{eq:interval_width}
\end{equation}
The interval characterization involves a tradeoff between reliability and sharpness. Although a wider interval can improve the probability of covering the measured path loss, an overly conservative interval weakens the usefulness of the prediction. Therefore, this work aims to obtain an accurate representative path loss prediction together with a reliable and compact prediction interval.

\section{Wireless Environment Knowledge Pool Construction} \label{sec:3}
\begin{figure}[!t]
\centerline{\includegraphics[width=\columnwidth]{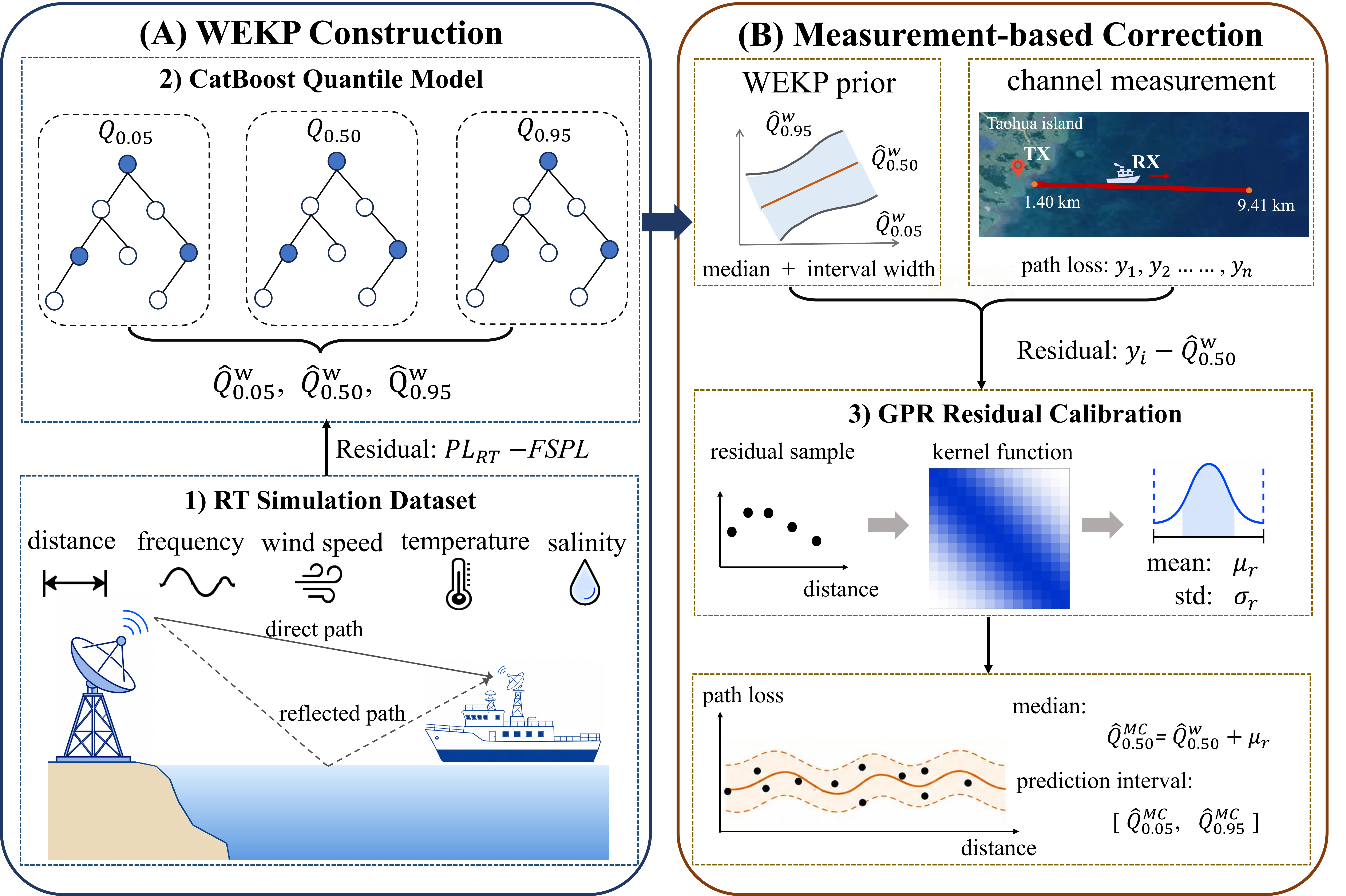}}
\caption{Overview of the proposed wireless environment knowledge pool-enhanced path loss prediction framework.}
\label{fig_main}
\end{figure}

This section presents the construction of the WEKP, as illustrated in Fig.~\ref{fig_main}.
The WEKP is built from RT simulation data to characterize the relationship between shore-to-ship propagation conditions and path loss quantiles. 
Specifically, it learns the mapping from environmental and system parameters to the quantiles of the residual between the RT-simulated path loss and the FSPL baseline.
The following subsections introduce the RT simulation dataset and the implementation method of the WEKP.

\subsection{Dataset Construction Based on Ray Tracing} \label{sub:3.1}
This subsection describes the RT simulation, which is to generate a large-scale dataset over a comprehensive parameter space. High-precision environmental reconstruction and accurate electromagnetic properties of materials are key prerequisites for ensuring the reliability of the simulation results \cite{jiaxin_2024}. Therefore, before introducing the mechanisms of radio wave propagation, we first present the methods for ocean wave modeling and seawater dielectric constant calculation. These methods enable accurate characterization of radio propagation, which is critical for reliable RT simulation in maritime environment.

The ocean wave spectrum is a common approach in wave modeling, which is typically derived from long-term sea surface observations by the researchers. Random sea surfaces are commonly generated using two approaches, the linear superposition method \cite{toporkov_2000} and the linear filtering method \cite{jiang_2015}. The linear filtering method uses the inverse fast Fourier transform (IFFT), which can generate rapid simulation of wave characteristics, and is therefore adopted in this paper. Various wave spectrum can be used to describe the sea surface characteristics, among which the PM wave spectrum is a widely used model. The PM wave spectrum is adopted to simulate sea surface conditions at different wind speeds, providing the environmental inputs for subsequent RT simulations. The simulated sea surface inevitably differs from the real sea surface to some extent. We focus on simulating wave characteristics under different wind speeds, rather than replicating the precise wave patterns in the measured environment. The formula for the PM wave spectrum is presented in \eqref{eq:2}.

\begin{equation}
S(\omega) = \frac{8.1 \times 10^{-3} \, g^2}{\omega^5} \exp\left( -\frac{0.74 \, g^4}{\omega^4 U^4} \right)\label{eq:2}
\end{equation}
where $\omega$ is the wave frequency, $g$ represents the gravitational acceleration, and $U$ is the wind speed at 19.5 $\mathrm{m}$ above the sea surface \cite{pierson_1964}.

The dielectric constant of seawater is often calculated using the Double Debye equation. This equation takes salinity, temperature, and frequency as the input and has the advantages of a wide frequency range and high precision \cite{Meissner_2004}.
\begin{equation}
\begin{split}
\varepsilon(T,S,f) = \frac{\varepsilon_s(T,S) - \varepsilon_1(T,S)}{1 + \mathrm{i}f/f_1(T,S)} + \frac{\varepsilon_1(T,S) - \varepsilon_{\infty}(T,S)}{1 + \mathrm{i}f/f_2(T,S)} 
\\+ \varepsilon_{\infty}(T,S) - \frac{\mathrm{i}\sigma(T,S)}{(2\pi\varepsilon_0)f} 
\label{eq:3}
\end{split}
\end{equation}
where $T$ is the seawater temperature, $S$ is the salinity, and $f$ is the frequency. The remaining parameters are defined in \cite{Meissner_2004}.

The undulation of the sea surface significantly affects the reflection characteristics. As sea surface roughness increases, the specular reflection weakens and the energy becomes more diffuse. To accurately represent this effect, a roughness coefficient is introduced to modify the Fresnel reflection coefficient calculation.
The Fresnel reflection coefficient and the roughness coefficient are given by \eqref{eq:4} and \eqref{eq:5}.
\begin{equation}
\begin{split}
R_{\perp} &= \frac{\cos\theta_i - \sqrt{\varepsilon_r - \sin^2\theta_i}}{\cos\theta_i + \sqrt{\varepsilon_r - \sin^2\theta_i}}
\\R_{\parallel} &= \frac{\varepsilon_r \cos\theta_i - \sqrt{\varepsilon_r - \sin^2\theta_i}}{\varepsilon_r \cos\theta_i + \sqrt{\varepsilon_r - \sin^2\theta_i}}
\label{eq:4}
\end{split}
\end{equation}
where $\theta_i$ denotes the incident angle, and $\varepsilon_r$ represents the relative permittivity of the sea surface.
  
\begin{equation}
\rho_s = \exp\left[-2\left( \frac{2\pi\sigma_h \sin\Psi}{\lambda} \right)^2 \right] \cdot I_0\left[ 2\left( \frac{2\pi\sigma_h \sin\Psi}{\lambda} \right)^2 \right]  
\label{eq:5}
\end{equation}
where \( \rho_s \) is the roughness coefficient, \( \sigma_h \) is the root mean square height, \( \Psi \) is the grazing angle of the incident ray, \( \lambda \) is the wavelength, and \( I_0(\cdot) \) is the zero-order modified Bessel function \cite{huang_2016}.

In the RT simulation, the direct and the first-order sea surface reflected paths are considered as the main propagation components. Using this simplified RT simulation setup, a simulation dataset is generated by varying maritime environmental and system parameters.
Specifically, the wind speed varies from 2 to 8 $\mathrm{m/s}$ with a step of 1 $\mathrm{m/s}$, while the frequency ranges from 2 to 14 $\mathrm{GHz}$ at 2 $\mathrm{GHz}$ intervals. Regarding the antenna configurations, six representative height combinations are selected to cover common shore-to-ship communication deployments. These combinations are, $(h_{\mathrm{TX}}, h_{\mathrm{RX}})=(10~\mathrm{m}, 5~\mathrm{m})$, $(10~\mathrm{m}, 10~\mathrm{m})$, $(15~\mathrm{m}, 5~\mathrm{m})$, $(20~\mathrm{m}, 5~\mathrm{m})$, $(30~\mathrm{m}, 10~\mathrm{m})$, and $(30~\mathrm{m}, 5~\mathrm{m})$. The horizontal distance extends from 200 to 2000 $\mathrm{m}$ with 200 $\mathrm{m}$ increments. In addition, the seawater temperature and salinity are set from 14 to 22 $^\circ\mathrm{C}$ in 2 $^\circ\mathrm{C}$ increments and 28 to 34 parts per thousand ($\mathrm{ppt}$) in 2 $\mathrm{ppt}$ increments, respectively. Therefore, the dataset comprises 58,800 unique samples, each consisting of a specific input parameter set and the total path loss obtained from the combination of the direct and the reflected paths. The measurement campaign described in Section \ref{sub:4.1} covers the distances up to 9.41 $\mathrm{km}$, whereas the RT simulations are conducted over the distances up to approximately 2 $\mathrm{km}$.
This difference is mainly due to the computational burden of maritime RT simulations. Unlike urban RT simulations, whose complexity is largely determined by the number of buildings, maritime RT simulations require discretization of the continuous sea surface. As the simulation distance increases, the number of sea surface grids grows rapidly, leading to high computational complexity.

\subsection{WEKP Implementation Using CatBoost Quantile Regression} \label{sub:3.2}

To construct the WEKP, the FSPL model is first adopted as a deterministic baseline. Rather than directly learning the RT-simulated path loss, the WEKP is formulated as a residual quantile mapping. Specifically, the WEKP learns the conditional quantiles of the residual between the RT-simulated path loss and the FSPL prediction from the input feature vector composed of environmental and system parameters defined in \eqref{eq:input_vector}. The RT residual path loss is defined as
\begin{equation}
\Delta PL_{\mathrm{RT}} = PL_{\mathrm{RT}} - \mathrm{FSPL}
\label{eq:rt_residual}
\end{equation}
where $\Delta PL_{\mathrm{RT}}$ denotes the residual between the RT-simulated path loss and the FSPL baseline, $PL_{\mathrm{RT}}$ denotes the path loss obtained from RT simulation, and $\mathrm{FSPL}$ denotes the FSPL baseline. 

CatBoost quantile regression is adopted as the implementation method of the WEKP. CatBoost is a gradient boosting decision tree model that constructs an ensemble of decision trees in a sequential manner to progressively reduce the loss left by the previous ensemble.
It utilizes oblivious trees as base predictors, which effectively mitigates overfitting and accelerates inference speed \cite{Prokhorenkova_2018}. In this paper, we employ CatBoost to effectively capture the complex nonlinear relationships between the input parameters and the RT residual path loss. The input feature of CatBoost is as shown in \eqref{eq:input_vector}, and the training target is $\Delta PL_{\mathrm{RT}}$. Instead of providing only a point prediction, CatBoost quantile regression estimates conditional quantiles of $\Delta PL_{\mathrm{RT}}$ at different probability levels.
For each quantile level $q$, the quantile prediction error is defined as
\begin{equation}
u_q = \Delta PL_{\mathrm{RT}} - \hat{Q}_{q}(\mathbf{x})
\label{eq:quantile_error}
\end{equation}
where $u_q$ denotes the error between the RT residual path loss and its estimated $q$-th conditional quantile, and $\hat{Q}_{q}(\mathbf{x})$ estimated $q$-th conditional quantile. The corresponding quantile loss is given by
\begin{equation}
\ell_q =
\max\left\{
-(1-q)u_q,\ q u_q
\right\}
\label{eq:quantile_loss}
\end{equation}
By minimizing this loss for different quantile levels, CatBoost learns multiple conditional quantiles of the RT residual path loss. The lower and upper quantiles describe the uncertainty range of the residual term, while the median quantile provides the residual point estimate.

After the residual quantiles are obtained, they are added to the FSPL baseline to construct the WEKP-based path loss quantiles. For a given quantile level $q$, the WEKP-based path loss quantile is expressed as
\begin{equation}
\hat{Q}^{\mathrm{w}}_{q}(\mathbf{x})=
\mathrm{FSPL} + \hat{Q}_{q}(\mathbf{x})
\label{eq:wekp_path_loss_quantile}
\end{equation}
where $\hat{Q}^{\mathrm{w}}_{q}(\mathbf{x})$ denotes the WEKP-based path loss quantile. In particular, $\hat{Q}^{\mathrm{w}}_{0.50}(\mathbf{x})$ is used as the WEKP-based median prediction.
For a prediction interval with a target coverage level of \(1-\alpha\), the WEKP-based prediction interval is constructed as
\begin{equation}
\hat{I}^{\mathrm{w}}_{1-\alpha}(\mathbf{x}) =
\left[
\hat{Q}^{\mathrm{w}}_{\alpha/2}(\mathbf{x}),
\hat{Q}^{\mathrm{w}}_{1-\alpha/2}(\mathbf{x})
\right]
\label{eq:wekp_interval}
\end{equation}
where $\hat{Q}^{\mathrm{w}}_{\alpha/2}(\mathbf{x})$ and $\hat{Q}^{\mathrm{w}}_{1-\alpha/2}(\mathbf{x})$ denote the lower and upper WEKP-based path loss quantiles, respectively. 
The WEKP provides the median path loss prediction and the corresponding prediction interval by learning the residual characteristics from RT simulation data. 
These predictions are further refined through measurement-based residual correction, resulting in the final path loss prediction and uncertainty characterization for practical shore-to-ship propagation scenarios.

\section{Measurement-based Residual Correction} \label{sec:4}
The WEKP constructed in Section~\ref{sec:3} provides the path loss quantiles from RT simulation data. In the RT simulation, the direct and first-order sea surface reflected paths are considered as the main propagation mechanisms. However, real shore-to-ship environment may include additional propagation effects that are not fully represented in the RT simulation. Therefore, when the WEKP is directly applied to the measurement scenario, a mismatch may exist between the WEKP-based prediction and the measured path loss.
Therefore, a small number of measurement samples are introduced for residual correction. Specifically, the residual between the measured path loss and the WEKP-based median prediction is modeled and used to refine the WEKP output. In the measurement-based residual correction stage, the WEKP prediction is regarded as the prior information. This correction enables the WEKP prior to better adapt to real shore-to-ship measurement conditions. The following subsections describe the measurement campaign and the measurement-based residual correction method.

\subsection{Measurement Campaign} \label{sub:4.1}
\begin{figure}[t]
\centering
\begin{subfigure}{0.9\columnwidth}
    \centering
    \includegraphics[width=\linewidth]{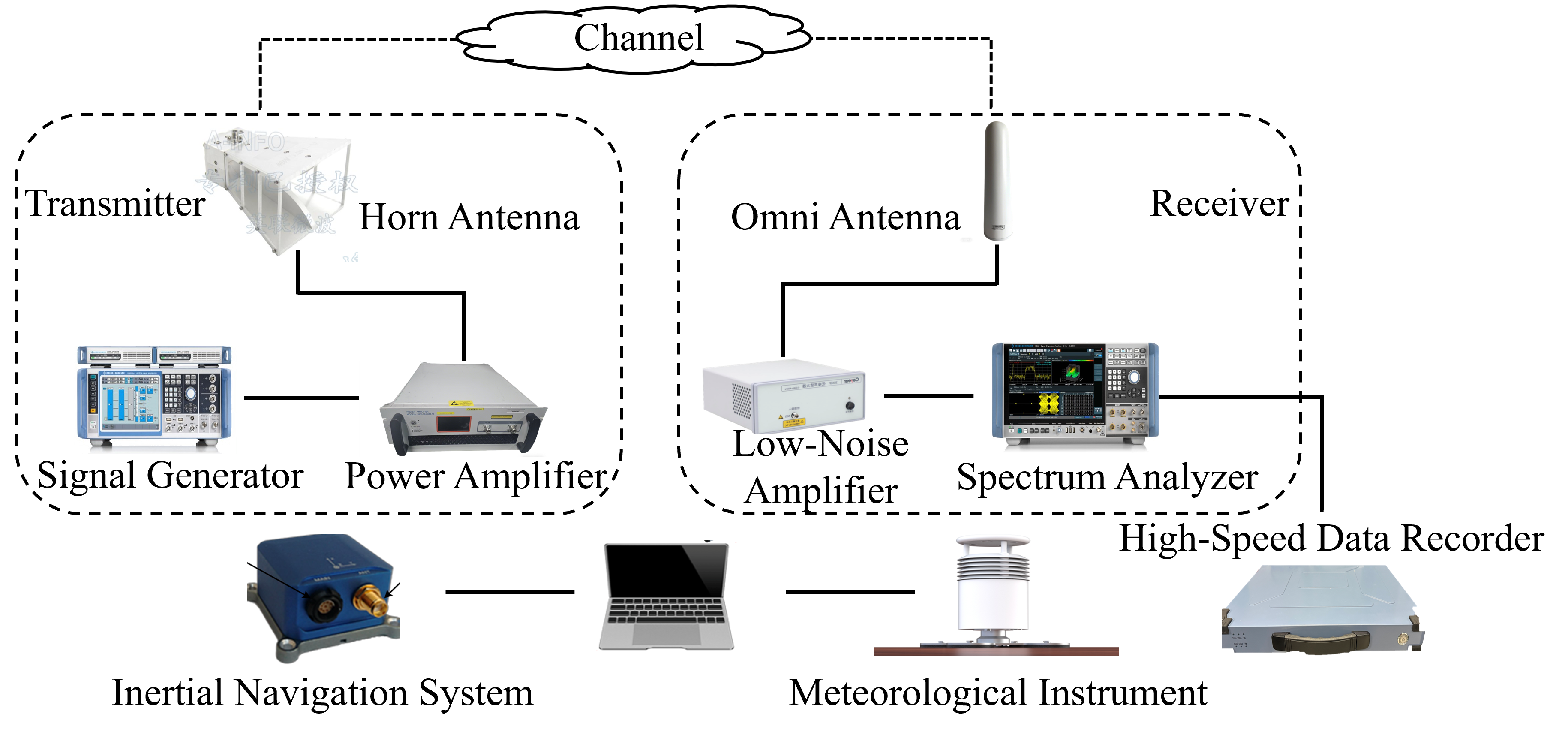}
    \caption{}
    \label{fig1}
\end{subfigure}
\begin{subfigure}{0.8\columnwidth}
    \centering
    \includegraphics[width=\linewidth]{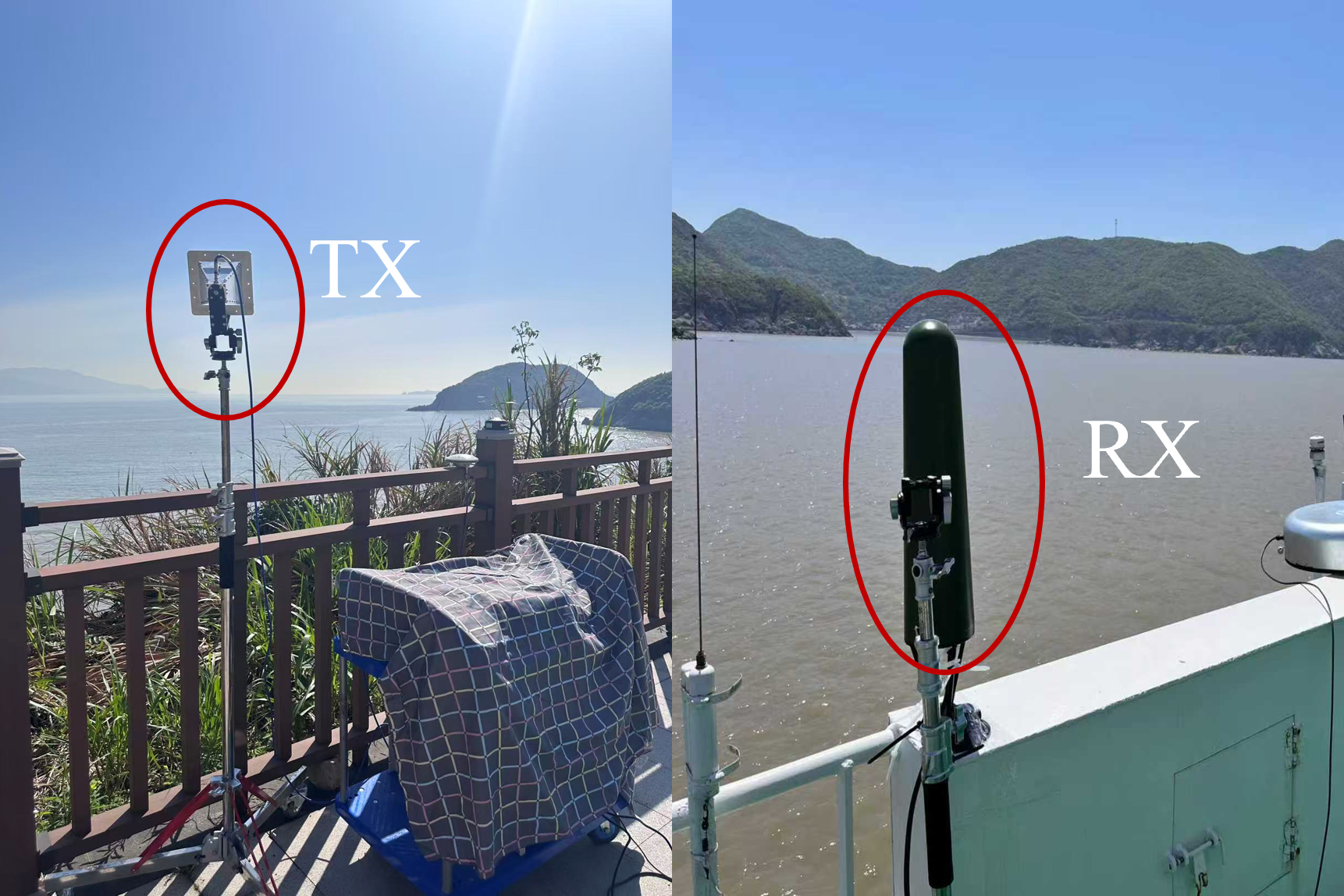}
    \caption{}
    \label{fig2}
\end{subfigure}
\begin{subfigure}{0.9\columnwidth}
    \centering
    \includegraphics[width=\linewidth]{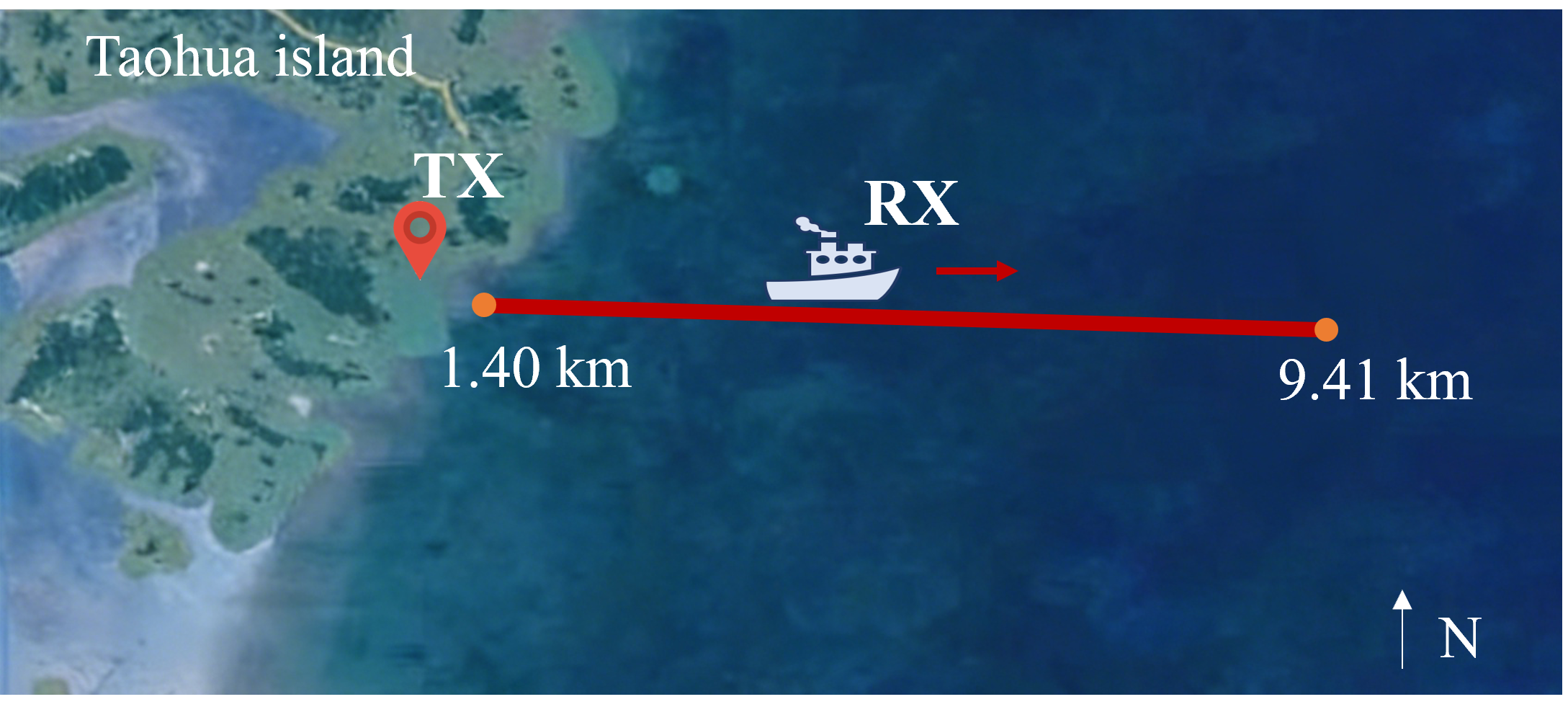}
    \caption{}
    \label{fig3}
\end{subfigure}
\caption{Measurement campaign in the shore-to-ship scenario. (a) Illustration of channel measurement platform. (b) Transmitter and receiver antennas for the shore-to-ship measurement. (c) Measurement route.}
\label{fig:sea}
\end{figure}

A shore-to-ship measurement campaign was conducted around Taohua Island, Zhoushan, Zhejiang Province, to collect real path loss data under practical maritime propagation conditions. The collected data are used for measurement-based residual correction and final performance evaluation.
The measurement platform comprises a vector signal generator (R\&S SMW200A), antennas, a power amplifier (MPA-30-6000-10), a low-noise amplifier (3840F), and a spectrum analyzer (R\&S FSW43) in Fig.~\ref{fig:sea}(\subref{fig1}). The GPS clock module provides a common timing reference to ensure synchronization between the transmitter and receiver units during the measurement campaign. At the transmitter, a horn antenna sends a pseudo-random code of length 2047, while an omnidirectional antenna at the receiver receives this signal as shown in Fig.~\ref{fig:sea}(\subref{fig2}). TABLE~\ref{tab:table1} lists the measurement center frequencies, bandwidths, and corresponding antenna gains. 

\begin{table}[!t]
    \centering 
    \caption{Channel Measurement Parameter Settings} 
    \label{tab:table1} 
    \renewcommand{\arraystretch}{1.1}
    \begin{tabular}{|c| c| c| c|}
        \hline
        \textbf{Center frequency (GHz)} & 4 & 4.5 & 5 \\
        \hline
        \textbf{TX gain (dBi)} & 11 & 12 & 11.5 \\
        \hline
        \textbf{RX gain (dBi)} & 2.13 & 1.56 & 2.26 \\
        \hline
        \textbf{Symbol rate (MHz/s)} & 100, 300 & 100, 300 & 100, 300 \\
        \hline
        \textbf{Bandwidth (MHz)} & 200, 600 & 200, 600 & 200, 600 \\
        \hline
    \end{tabular}
\end{table}

The TX antenna was set up on a flat terrain site on the Taohua Island at 29.7877$^\circ$N, 122.3096$^\circ$E. The RX antenna was fixed on the research ship $Zhe Yu Ke 2$. The ship was equipped with an inertial navigation system, which recorded real-time variations in the RX antenna height and attitude angles. A meteorological instrument was also placed on the ship, which recorded sea surface wind speed and other meteorological data simultaneously. During the measurement, the research ship moved along a preset course and covered a distance of 1.40–9.41 $\mathrm{km}$ as illustrated in Fig.~\ref{fig:sea}(\subref{fig3}). 
During the measurement campaign, the research ship recorded several key environmental parameters. The seawater temperature was between 15.3 and 16.0 $^\circ\mathrm{C}$. The salinity was between 32.4 and 34.6 $\mathrm{ppt}$. Additionally, the wind speed at a height of 10 $\mathrm{m}$ was recorded from 2.5 to 3.1 $\mathrm{m/s}$. 

In order to extract the real channel impulse response (CIR) from the measurement data, it is necessary to remove the impact of the measurement system hardware. We adopt a back-to-back (B2B) calibration method before conducting the measurement. Specifically, the TX and RX antennas are replaced with the attenuator of the known attenuation value, and the calibration signal is obtained through this configuration \cite{miao_2023}. The processing of the sampled IQ data proceeds as follows. First, the time-domain IQ data are transformed into the frequency domain through the fast Fourier transform. Subsequently, the transformed data are divided by the calibration signal to eliminate the influence of the system. Finally, the IFFT is applied to the ratio, and the time-domain CIR can be obtained. The path loss value can be written as
\begin{equation}
PL_{meas}[\text{dB}] = -10 \log_{10}\left(\frac{1}{N} \sum_{n=1}^{N} \sum_{m=1}^{M} \lvert h_{m,n}(t,\tau) \rvert^2 \right)
\label{eq:1}
\end{equation}
where $h_{m,n}(\tau)$ is the CIR sample at the $m$-th delay bin for the $n$-th measurement snapshot. The parameters $M$ and $N$ denote the numbers of delay bins and measurement snapshots, respectively.
 After data cleaning, 25,587 valid samples are retained across the three measured frequency points. 

\subsection{Gaussian Process Regression for Residual Correction} \label{sub:4.2}
When the WEKP prior is directly applied to real measurement scenarios, a mismatch may still exist between the WEKP-based prediction and the measured path loss. This mismatch is mainly caused by the difference between the RT simulation settings and the practical measurement environment. Therefore, a small number of measurement samples are introduced to perform measurement-based residual correction.

GPR is adopted to learn the residual between the measured path loss and the WEKP-based median prediction. The purpose of GPR is not to replace the WEKP prior or directly predict the measured path loss. Instead, GPR only models the remaining difference after the WEKP-based median prediction has been obtained. The residual is defined as
\begin{equation}
r =
PL_{\mathrm{meas}}
-\hat{Q}^{\mathrm{w}}_{0.50}(\mathbf{x})
\label{eq:measurement_residual}
\end{equation}
where $PL_{\mathrm{meas}}$ denotes the measured path loss, $\hat{Q}^{\mathrm{w}}_{0.50}$ denotes the WEKP-based median prediction, and $r$ denotes the corresponding residual. By learning this residual term, the correction model focuses on the mismatch between the WEKP prior and the real measurement, rather than directly learning the entire path loss relationship from measurement samples alone.

GPR is suitable for this residual correction task because it can model nonlinear correlations using a small number of measurement samples. In addition, GPR provides both the predicted mean and variance of the residual, which can be used for point correction and interval construction, respectively \cite{Williams_1995}. In the proposed framework, the input feature vector of GPR is denoted as $\mathbf{z}$. It is constructed by adding the original feature vector $\mathbf{x}$ with the WEKP-based interval width. For a target coverage level of $1-\alpha$, the WEKP-based interval width is defined as
\begin{equation}
W^{\mathrm{w}}(\mathbf{x}) = 
\hat{Q}^{\mathrm{w}}_{1-\alpha/2}(\mathbf{x})
-\hat{Q}^{\mathrm{w}}_{\alpha/2}(\mathbf{x}),
\label{eq:wekp_width}
\end{equation}
The GPR input feature vector is then given by
\begin{equation}
\mathbf{z} =
\left[
\mathbf{x},
W^{\mathrm{w}}(\mathbf{x})
\right]
\label{eq:gpr_input}
\end{equation}
The WEKP-based interval width provides additional uncertainty information under different environmental and system conditions. 
By incorporating $W^{\mathrm{w}}(\mathbf{x})$ as an input feature, GPR can utilize this prior uncertainty information to better characterize the variation of residuals between WEKP-based predictions and measurement results under different propagation conditions.
Assume that $N_{\mathrm{m}}$ measurement samples are selected for measurement-based residual correction. For the $i$-th selected sample, the input feature vector of the GPR is $\mathbf{z}_i$, and the corresponding regression target is the measurement residual $r_i$. Therefore, the training set for GPR is defined as
\begin{equation}
\mathcal{D}_{\mathrm{i}} =
\left\{
\left(
\mathbf{z}_i,
r_i
\right)
\right\}_{i=1}^{N_{\mathrm{m}}}
\label{eq:gpr_training_set}
\end{equation}
where $\mathbf{z}_i$ denotes the input feature vector of the $i$-th sample for residual correction, and $r_i$ denotes the corresponding residual between the measured path loss and the WEKP-based median prediction.

The residual correction function is modeled as a Gaussian process:
\begin{equation}
r(\mathbf{z}) \sim
\mathcal{GP}
\left(
0,
k(\mathbf{z},\mathbf{z}')
\right)
\label{eq:gpr_prior}
\end{equation}
where \(k(\mathbf{z},\mathbf{z}')\) is the kernel function that defines the covariance between residual function values at different input feature vectors \(\mathbf{z}\) and \(\mathbf{z}'\). A larger kernel value indicates stronger covariance between the residuals associated with the two input feature vectors \cite{Jang_2022}. Using the selected correction samples, GPR estimates the posterior predictive distribution of the residual correction for a new input $\mathbf{z}$. The predictive mean and variance of the residual correction are denoted as $\mu_r(\mathbf{z})$ and $\sigma_r^2(\mathbf{z})$, respectively. The residual mean is used to correct the WEKP-based median prediction, while the residual variance is used to construct the corrected prediction interval.
The corrected median prediction is defined as
\begin{equation}
\hat{Q}^{\mathrm{mc}}_{0.50}(\mathbf{z})
=\hat{Q}^{\mathrm{w}}_{0.50}(\mathbf{x})
+
\mu_r(\mathbf{z})
\label{eq:mc_median}
\end{equation}
where $\hat{Q}^{\mathrm{mc}}_{0.50}(\mathbf{z})$ denotes the corrected median path loss prediction. 
The corrected prediction interval is constructed by using the uncertainty information provided by the GPR residual correction. For a target coverage level of $1-\alpha$, the interval is defined as
\begin{equation}
\hat{I}^{\mathrm{mc}}_{1-\alpha}(\mathbf{z})
=\left[
\hat{Q}^{\mathrm{mc}}_{0.50}(\mathbf{z})-k \sigma_r(\mathbf{z}),
\
\hat{Q}^{\mathrm{mc}}_{0.50}(\mathbf{z})
+
k \sigma_r(\mathbf{z})
\right]
\label{eq:mc_interval}
\end{equation}
where $k$ is a fixed scaling factor determined by the target coverage level. 
For a target coverage level of 90\%, $k$ is set to 1.645. 
Through this residual correction process, the proposed framework adapts the WEKP to the real measurement scenarios by learning the discrepancy between simulation and measurement data. The resulting prediction incorporates both WEKP prior and measurement information, providing improved path loss estimation with uncertainty characterization.

\section{Experimental Results and Analysis} \label{sec:5}

This section presents the performance evaluation of the proposed WEKP-PLP framework for shore-to-ship path loss prediction and uncertainty characterization. The experiments include evaluations based on RT simulation data and measurement data, where the former is used to construct the WEKP prior and the latter is used for residual correction and validation. 
Since the proposed framework aims to provide both point prediction and interval characterization, the evaluation does not only focus on MAE, RMSE, and $R^2$. The metrics related to interval characterization, including coverage, interval width, and Winkler score, are also considered. Among these metrics, the Winkler score is particularly important because it jointly evaluates interval sharpness and the penalty for observations falling outside the prediction interval.

\subsection{Performance Analysis of WEKP Construction}

This subsection evaluates the construction of the WEKP based on the RT simulation dataset. Specifically, the WEKP is designed to learn the mapping from environmental and system parameters to the residual quantiles between the RT-simulated path loss and the FSPL baseline. Since different regression models can be used to approximate this mapping, four representative models are considered as candidate implementations of the WEKP, including Extreme Gradient Boosting (XGBoost), CatBoost, Natural Gradient Boosting (NGBoost), and multilayer perceptron (MLP). The RT simulation dataset is randomly divided into training, validation, and test sets with a ratio of 8:1:1, and Optuna is used to search for the optimal hyperparameter combination of each model within the predefined parameter ranges. All candidate models are trained using the feature configuration, which includes frequency, horizontal distance, seawater temperature, salinity, and wind speed. 
The target coverage level is set to 90\%. Accordingly, the 0.05, 0.50, and 0.95 quantiles are predicted to obtain the lower bound, median prediction, and upper bound, respectively.

The candidate models are first compared in terms of point prediction performance. The results indicate that all evaluated models obtain similar point prediction accuracy, with RMSE values around 1.84 dB and $R^2$ values around 0.95. These results suggest that learning the residual component simplifies the prediction task compared with directly modeling the entire path loss.
The FSPL baseline represents the main attenuation trend caused by propagation distance and carrier frequency, while the candidate models focus on learning the residual component.
Therefore, point prediction metrics alone are not sufficient for selecting the implementation of the WEKP.
Since the proposed framework also needs to characterize path loss uncertainty, interval prediction performance is further considered. Coverage indicates whether the prediction interval includes the true path loss with a probability close to the target level. Mean width indicates the compactness of the interval. However, these two metrics should not be interpreted separately, because a wider interval may increase coverage but reduce practical usefulness. Therefore, the Winkler score is used as the main criterion for model selection, since it penalizes both excessive interval width and missed observations. 
The Winkler score is defined as
\begin{equation}
\begin{aligned}
W ={}& \left(\hat{Q}_{1-\alpha/2}-\hat{Q}_{\alpha/2}\right) \\
&+ \frac{2}{\alpha}
\left(\hat{Q}_{\alpha/2}-y\right)
\mathbb{I}\left(y<\hat{Q}_{\alpha/2}\right) \\
&+ \frac{2}{\alpha}
\left(y-\hat{Q}_{1-\alpha/2}\right)
\mathbb{I}\left(y>\hat{Q}_{1-\alpha/2}\right).
\end{aligned}
\end{equation}
where $\hat{Q}_{\alpha/2}$ and $\hat{Q}_{1-\alpha/2}$ denote the estimated lower and upper quantiles, respectively, $y$ represents the ground truth path loss, and $\alpha$ is the significance level. $\mathbb{I}(\cdot)$ is the indicator function, which equals 1 when the condition inside the parentheses is true and 0 otherwise. A lower Winkler score indicates a better balance between interval sharpness and coverage reliability.

As shown in Table~\ref{tab:table2}, the evaluated models show the differences in the performance of interval characterization.
For the MLP, the training objective has already considered the pinball loss, interval width, and median prediction accuracy. 
However, the MLP still produces overly narrow intervals, with a mean width of only 0.89 dB and a maximum width of 3.73 dB. 
Its coverage is only 68.05\%, which is much lower than the target coverage level. 
In contrast, the gradient boosting models provide more reliable interval estimates. 
The XGBoost, CatBoost, and NGBoost all achieve the target level of 90\%, while maintaining moderate interval widths. 
Among them, the CatBoost obtains a coverage of 90.03\%, a mean interval width of 4.72 dB, and the lowest Winkler score of 7.28. 
This result suggests that the CatBoost achieves the better balance between interval reliability and sharpness.
Therefore, CatBoost is selected as the implementation of the WEKP in the following experiments. 
The final CatBoost hyperparameters are set as follows, iterations of 1000, learning rate of 0.04, tree depth of 6, L2 leaf regularization of 1.89, and minimum data in leaf of 80.
The median quantile provides the WEKP-based point prior, while the lower and upper quantiles construct the WEKP-based interval prior.

\begin{table}[!t]
\centering
\caption{Performance comparison of different models on the RT simulation dataset}
\label{tab:table2}
\renewcommand{\arraystretch}{1.1}
\begin{tabular}{|c|c|c|c|c|c|c|c|}
\hline
\textbf{Model} 
& \makecell{\textbf{Coverage}} 
& \makecell{\textbf{Mean Width}\\\textbf{(dB)}} 
& \makecell{\textbf{Max Width}\\\textbf{(dB)}} 
& \makecell{\textbf{Winkler}\\\textbf{Score}} \\
\hline
XGBoost & 90.56\% & 4.66  & 11.22 & 7.52 \\
\hline
CatBoost &  90.03\% & 4.72  & 13.27 & 7.28 \\
\hline
NGBoost &  90.36\% & 4.73  & 15.66 & 7.38 \\
\hline
MLP  & 68.05\% & 0.89  & 3.73 & 11.83 \\
\hline
\end{tabular}
\end{table}

\begin{figure}[!t]
\centerline{\includegraphics[width=\columnwidth]{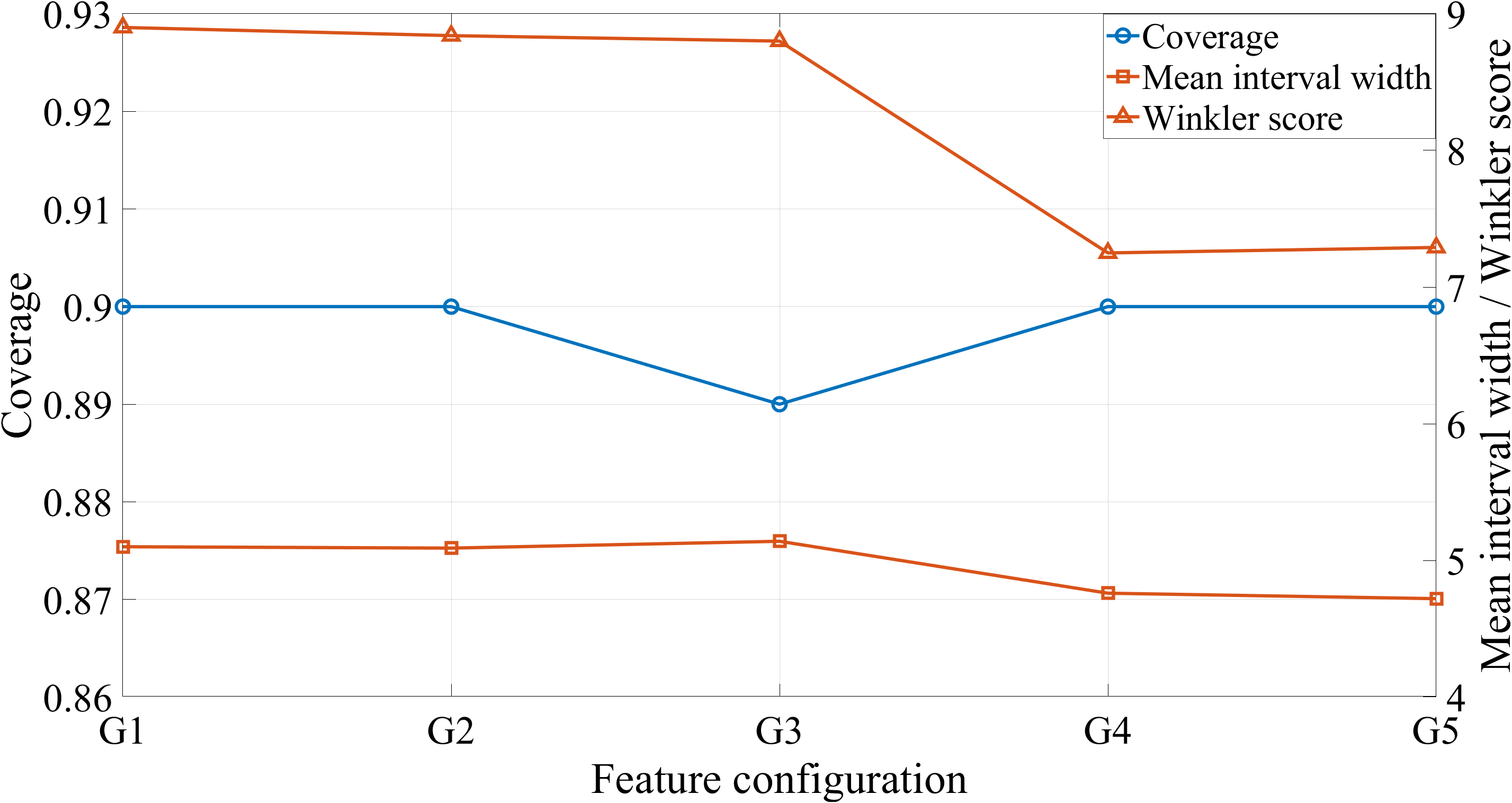}}
\caption{Feature ablation results under different input configurations.}
\label{fig:feature}
\end{figure}

To investigate the influence of different input features on the construction of the WEKP, a feature ablation study is conducted using the RT simulation dataset.
Figure~\ref{fig:feature} presents the performance comparison under different input feature configurations.
The five feature groups are considered to evaluate the contribution of system, environmental, and dielectric parameters. 
G1 feature set uses horizontal distance and frequency as the input features, and the residual target is constructed with respect to the 2D FSPL baseline. Here, 2D FSPL refers to the FSPL baseline calculated using the horizontal propagation distance. 
G2 feature set uses 3D distance and frequency as the input features, and the residual target is constructed using the FSPL baseline calculated from the three-dimensional distance between the transmitter and the receiver.
G3 feature set adds the antenna height difference to the G1, while still using the 2D FSPL baseline. 
G4 feature set further introduces environmental parameters, including wind speed, seawater temperature, and salinity, together with horizontal distance and frequency. 
G5 feature set additionally includes the real and imaginary parts of the seawater dielectric constant and the conductivity based on G4.
The comparison between G1 and G2 shows only negligible performance differences, indicating that the choice between 2D and 3D distance, both as an input feature and for FSPL calculation, has little impact on the prediction results.
The comparison between G1 and G3 shows only marginal improvement after introducing the antenna height difference. 
Therefore, the antenna height difference is excluded from the final input features because its contribution is limited, while its real-time acquisition requires additional sensing equipment.
Compared with the features in G1, the introduction of wind speed, temperature, and salinity in G4 changes the interval prediction behavior. 
This suggests that these environmental variables mainly contribute to the modeling of RT residual uncertainty. 
Furthermore, the improvement from G4 to G5 is limited, indicating that the dielectric variables provide only marginal additional information because they are derived from frequency, temperature, and salinity.

\begin{figure}[!t]
\centerline{\includegraphics[width=\columnwidth]{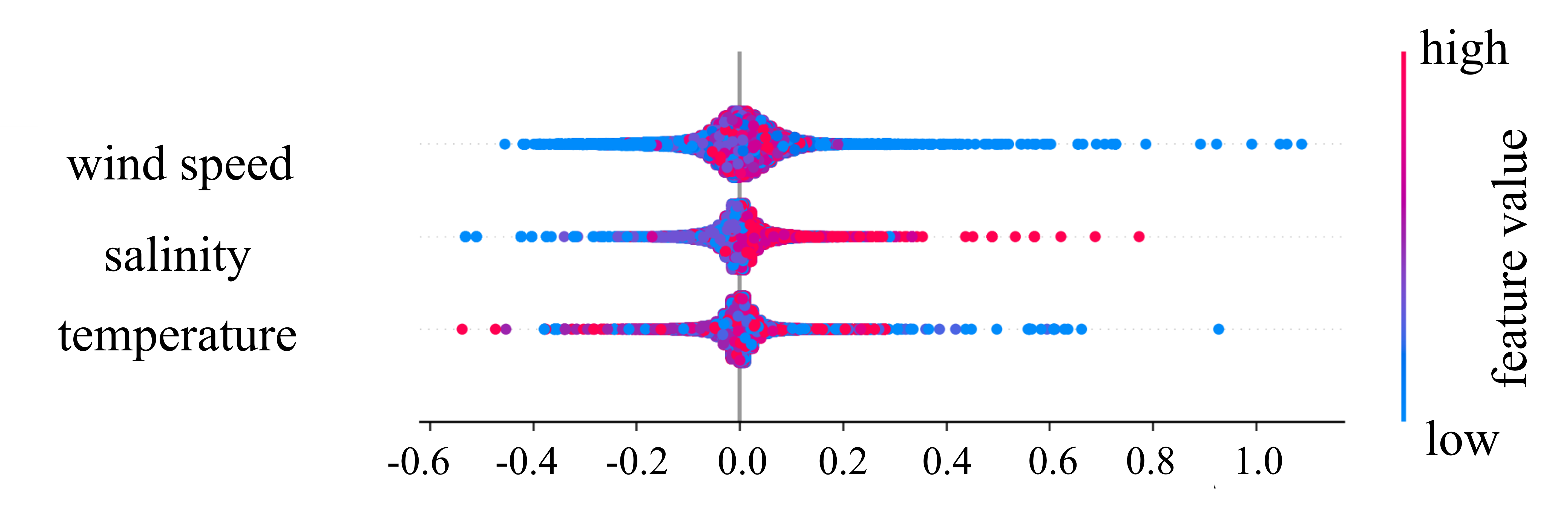}}
\caption{SHAP value distribution of environmental parameters.}
\label{fig:shap}
\end{figure}

To further interpret the contribution of environmental parameters, SHAP analysis is performed \cite{Lundberg_2017,sun_2022}. 
As shown in Fig.~\ref{fig:shap}, wind speed shows a wider SHAP value distribution among the environmental variables, indicating that it plays an important role in the learned RT residual variation. 
The SHAP values of wind speed are distributed on both positive and negative regions, which suggests that its effect on the RT residual is not purely monotonic.
This behavior is consistent with the influence of wind speed on sea surface reflection. 
As the wind speed increases, the sea surface becomes rougher, and the reflected components from the sea surface tends to decrease. 
Consequently, the interference between the direct path and the sea reflected components becomes weaker, which can reduce path loss fluctuations. 
In contrast, under lower wind speed conditions, the sea surface is closer to a smooth reflecting surface, and the reflected components may retain stronger coherent energy. 
The stronger reflected components can have a greater impact on the superposition with the direct path at the receiver, leading to larger variations in the RT residual.
The SHAP distributions of salinity and temperature also extend to both positive and negative regions, but their ranges are smaller than that of wind speed. 
This indicates that salinity and temperature also affect the learned residual behavior, mainly through their influence on the seawater dielectric constant and reflection coefficient. 
Compared with wind speed, their effects on the reflected components are relatively weaker in the considered RT simulation dataset. 

Overall, this subsection demonstrates the feasibility of constructing the WEKP from RT simulation data. 
The model comparison shows that CatBoost provides the balance between interval reliability and sharpness. 
The input feature configuration evaluation and SHAP analyses further confirm that environmental parameters contribute mainly to RT residual uncertainty characterization. 
Therefore, the constructed WEKP can provide both point prediction and interval information for the following measurement-based residual correction.

\subsection{Interval Behavior and Generalization Analysis of WEKP}

\begin{figure}[!t]
\centerline{\includegraphics[width=\columnwidth]{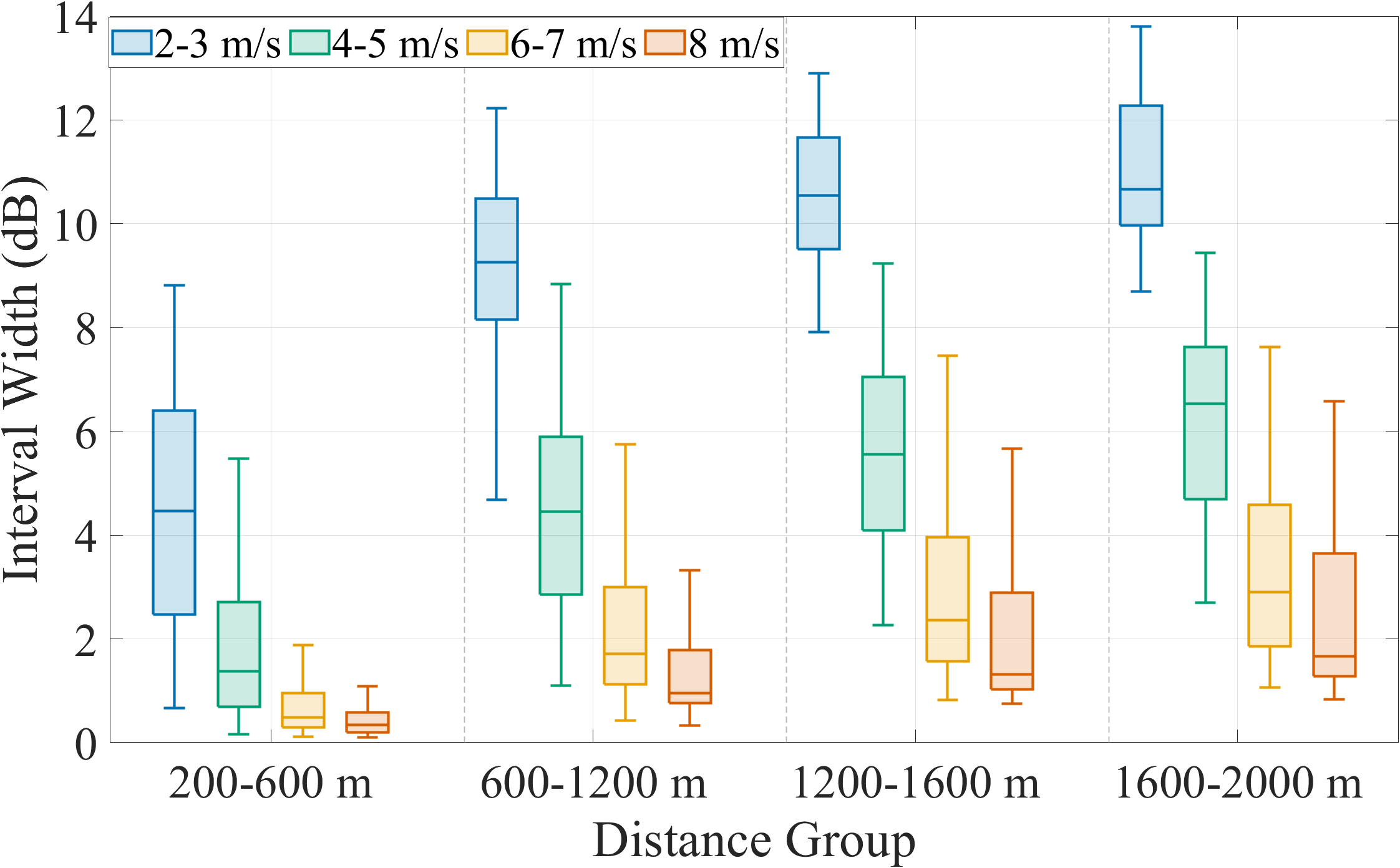}}
\caption{Interval width variation under different distance and wind speed conditions.}
\label{fig:width}
\end{figure}

This subsection evaluates whether the WEKP can effectively replace the RT simulation when the parameters of practical propagation scenarios are provided. 
The purpose is to examine whether WEKP can reproduce the path loss prediction and uncertainty information obtained from RT simulation. 
Furthermore, the variation in interval width is analyzed to investigate whether WEKP can characterize prediction uncertainty under different propagation conditions.

An additional RT simulation is conducted using a subset of environmental and system parameters extracted from the measurement scenario. 
This setting is designed to evaluate whether the WEKP can approximate RT simulation results under practical scenario parameters without rerunning RT simulations for the measurement conditions.
Under this setting, the WEKP achieves an RMSE of 2.77 dB.
The coverage remains 91.70\%, meeting the target coverage level. 
Meanwhile, the mean interval width increases to 10.21 dB, and the maximum interval width reaches 11.94 dB. 
These results suggest that the WEKP still provides reliable interval prediction under practical measurement related parameters, although its point prediction accuracy is affected by the change in input parameter. 
The relatively large interval width further suggests that the uncertainty of RT simulation may vary under these practical scenario parameters. 
Therefore, it is necessary to further analyze the relationship between the interval width and key propagation conditions, such as propagation distance and wind speed, to provide possible explanations for the wider prediction intervals.

Figure~\ref{fig:width} shows the distribution of the prediction interval width on the RT simulation test set used for WEKP construction. 
The samples are grouped according to propagation distance and wind speed.
In each box, the lower and upper edges represent the 25th and 75th percentiles, respectively, and the central line represents the median value. 
Therefore, the box height reflects the dispersion of the prediction interval width within each group.
The results show that the prediction interval width is strongly related to both propagation distance and wind speed. 
For the low wind speed group, the interval width increases significantly as the propagation distance becomes larger. 
This suggests that long propagation distance under relatively smooth sea surface conditions introduces larger uncertainty in RT residual prediction.
In addition, under the same distance group, the interval width generally decreases as the wind speed increases. 
This behavior is reasonable because higher wind speed increases sea surface roughness and weakens the reflected components from the sea surface. 
As a result, the interference between the direct path and the sea reflected components becomes weaker, leading to smaller path loss fluctuations caused by coherent superposition.
These results provide a possible explanation for the wider prediction intervals observed under related parameters of practical measurement. 
They also indicate that the WEKP provides adaptive prediction intervals rather than a fixed uncertainty range, with the interval varying according to the propagation distance and wind speed.

Overall, these results suggest that the WEKP can provide uncertainty information under different propagation conditions. 
In the additional RT simulation based on selected measurement scenario parameters, the point prediction accuracy moderately decreases, but the coverage still satisfies the target coverage requirement. 
The interval width analysis further suggests that the predicted uncertainty is related to propagation distance and wind speed. 
These results indicate that the WEKP is capable of providing adaptive uncertainty estimates under different propagation conditions.

\subsection{Measurement-based Residual Correction}

\begin{figure*}[!t]
    \centering
    \begin{subfigure}{0.31\linewidth}
        \centering
        \includegraphics[width=\linewidth]{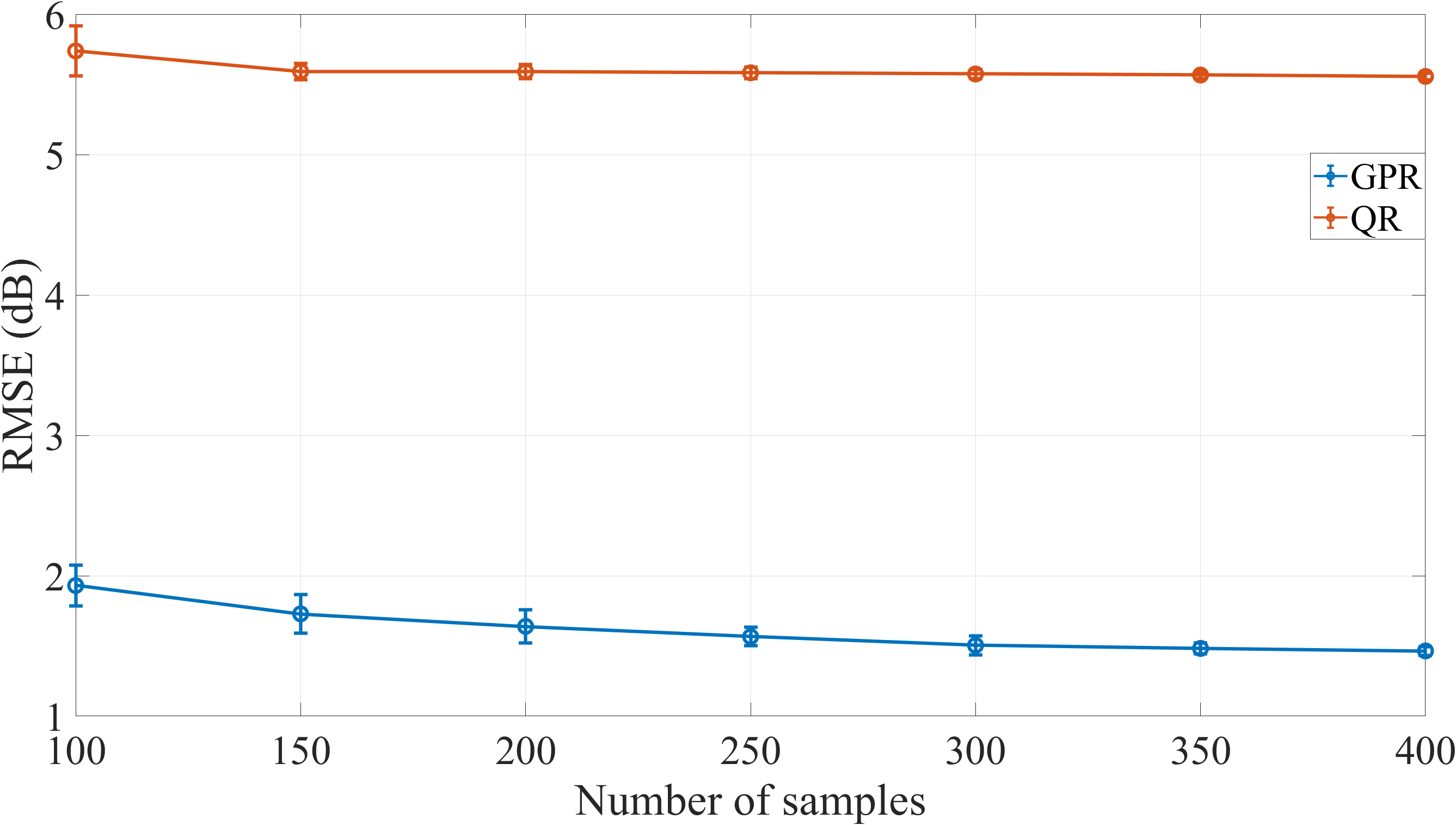} 
        \caption{}
    \end{subfigure}
    \hfill
    \begin{subfigure}{0.31\linewidth}
        \centering
        \includegraphics[width=\linewidth]{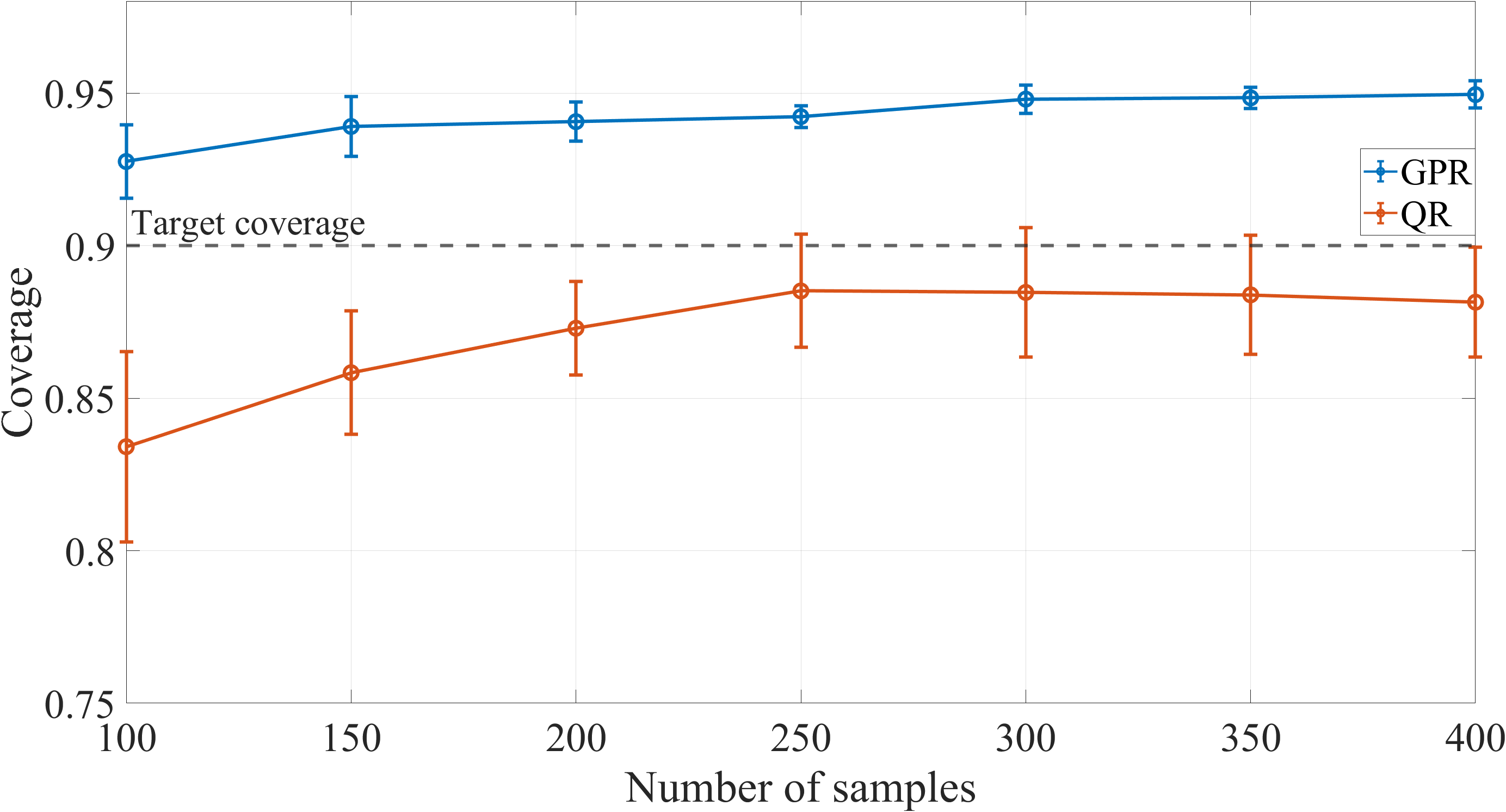} 
        \caption{}
    \end{subfigure}
    \hfill
    \begin{subfigure}{0.31\linewidth}
        \centering
        \includegraphics[width=\linewidth]{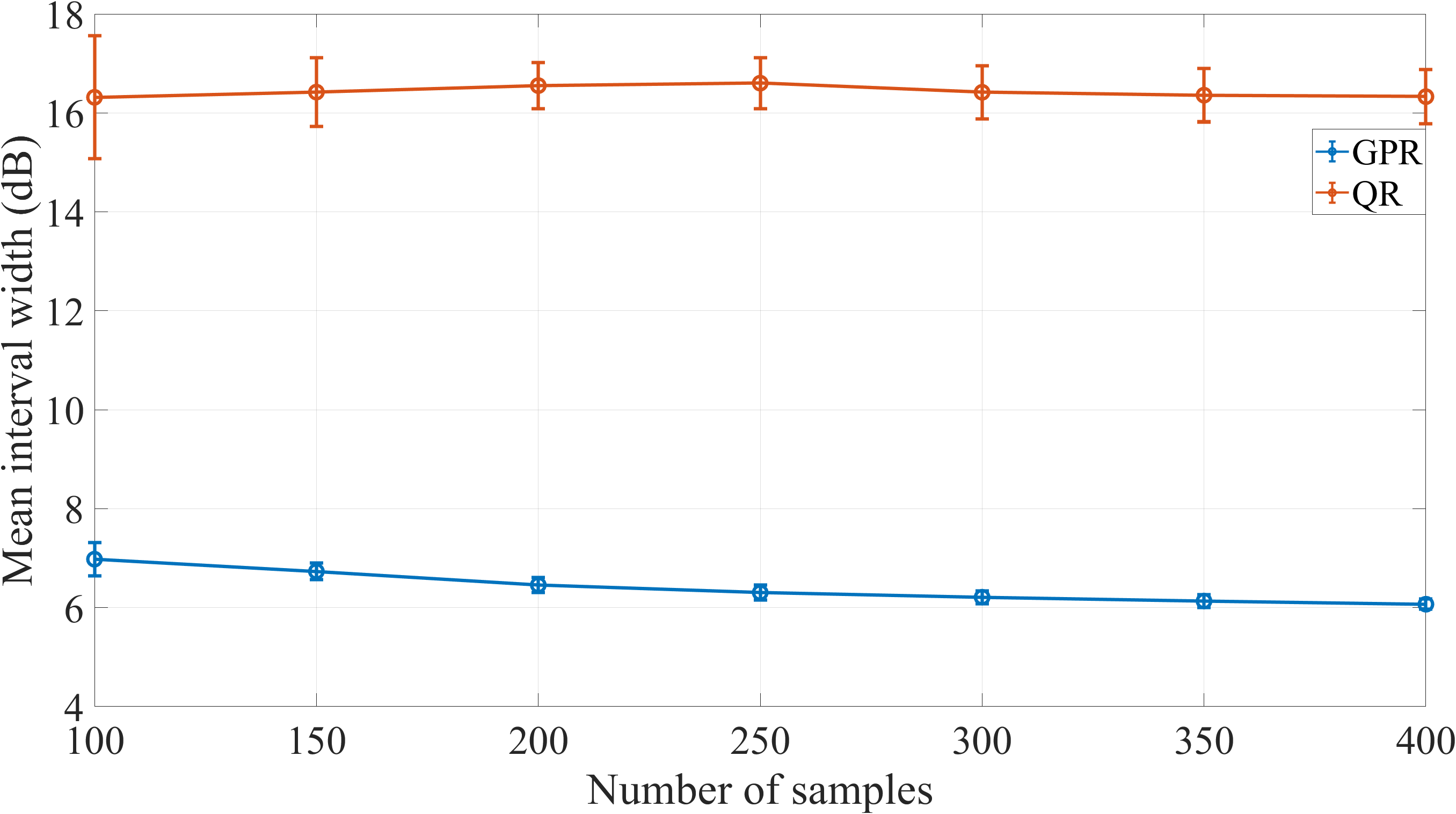} 
        \caption{} 
    \end{subfigure}
    \caption{Performance comparison between GPR and QR under different numbers of correction samples. (a) RMSE. (b) Coverage. (c)  Mean width.}
    \label{fig:number}
\end{figure*}

\begin{figure*}[!t]
    \centering
    \begin{subfigure}{0.31\linewidth}
        \centering
        \includegraphics[width=\linewidth]{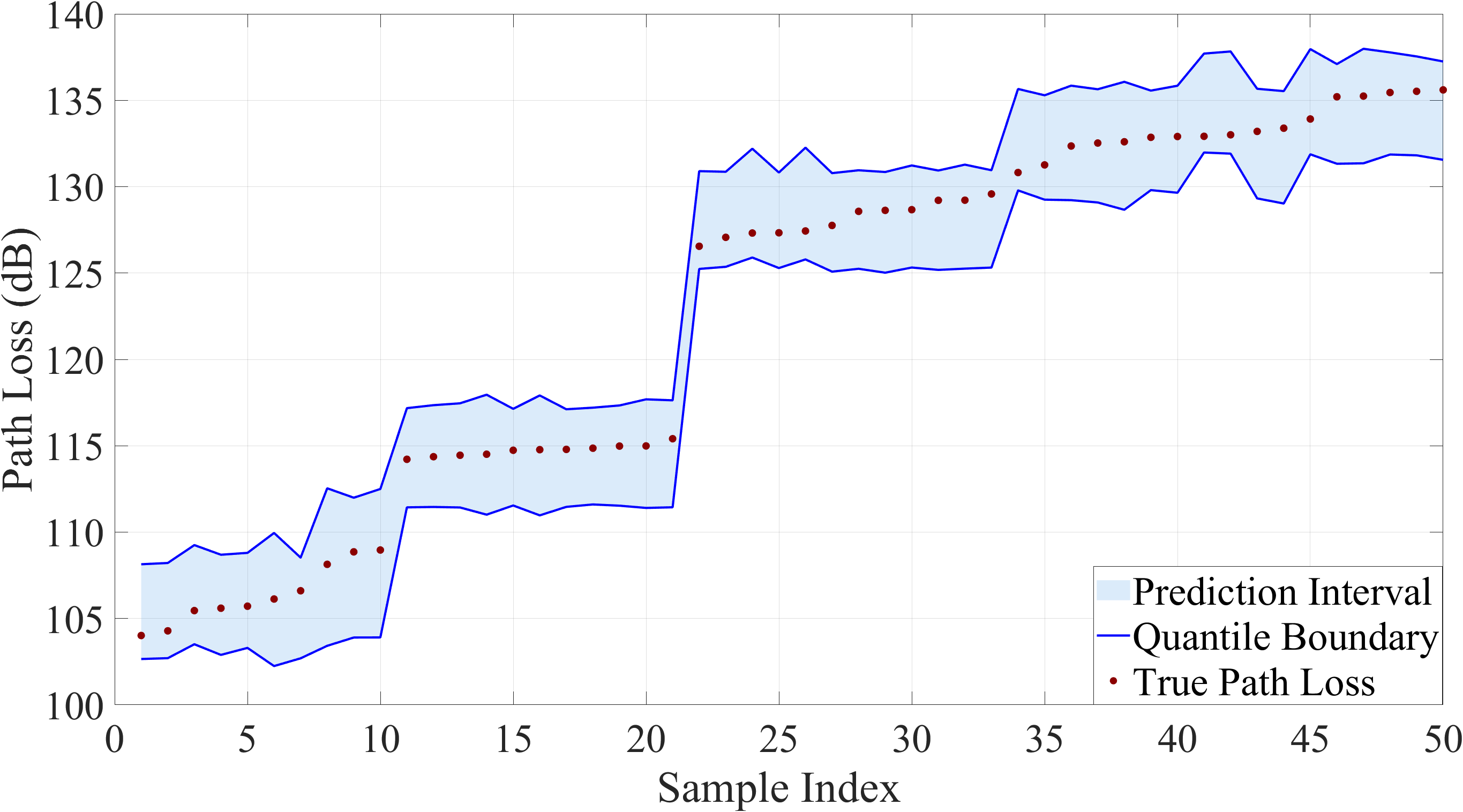} 
        \caption{}
    \end{subfigure}
    \hfill
    \begin{subfigure}{0.31\linewidth}
        \centering
        \includegraphics[width=\linewidth]{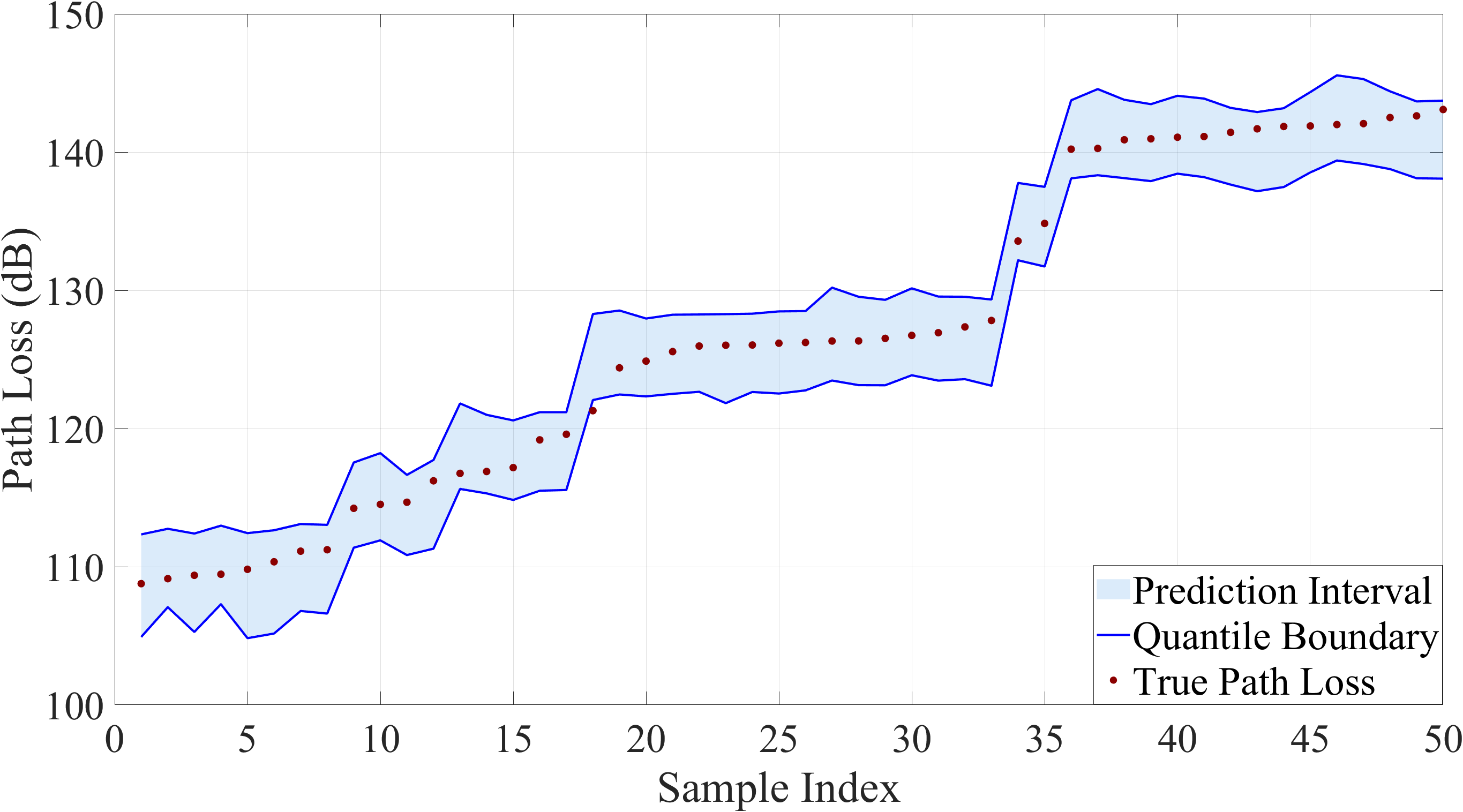} 
        \caption{}
    \end{subfigure}
    \hfill
    \begin{subfigure}{0.31\linewidth}
        \centering
        \includegraphics[width=\linewidth]{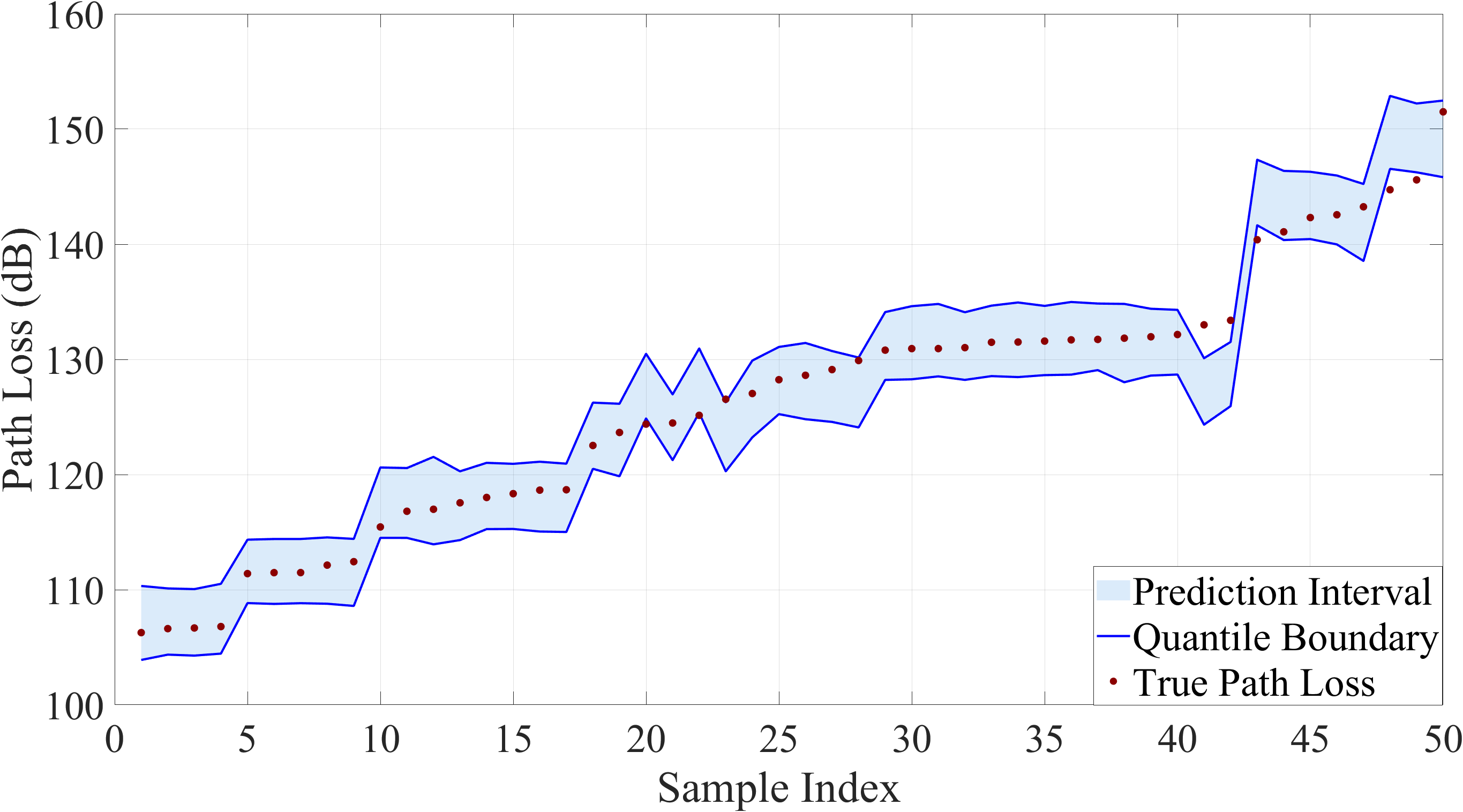} 
        \caption{} 
    \end{subfigure}
    \caption{Prediction intervals of the proposed model at different frequencies. (a) 4GHz. (b) 4.5GHz. (c) 5GHz.}
    \label{fig:interval}
\end{figure*}

\begin{figure}[t]
\centering
\includegraphics[width=0.48\textwidth]{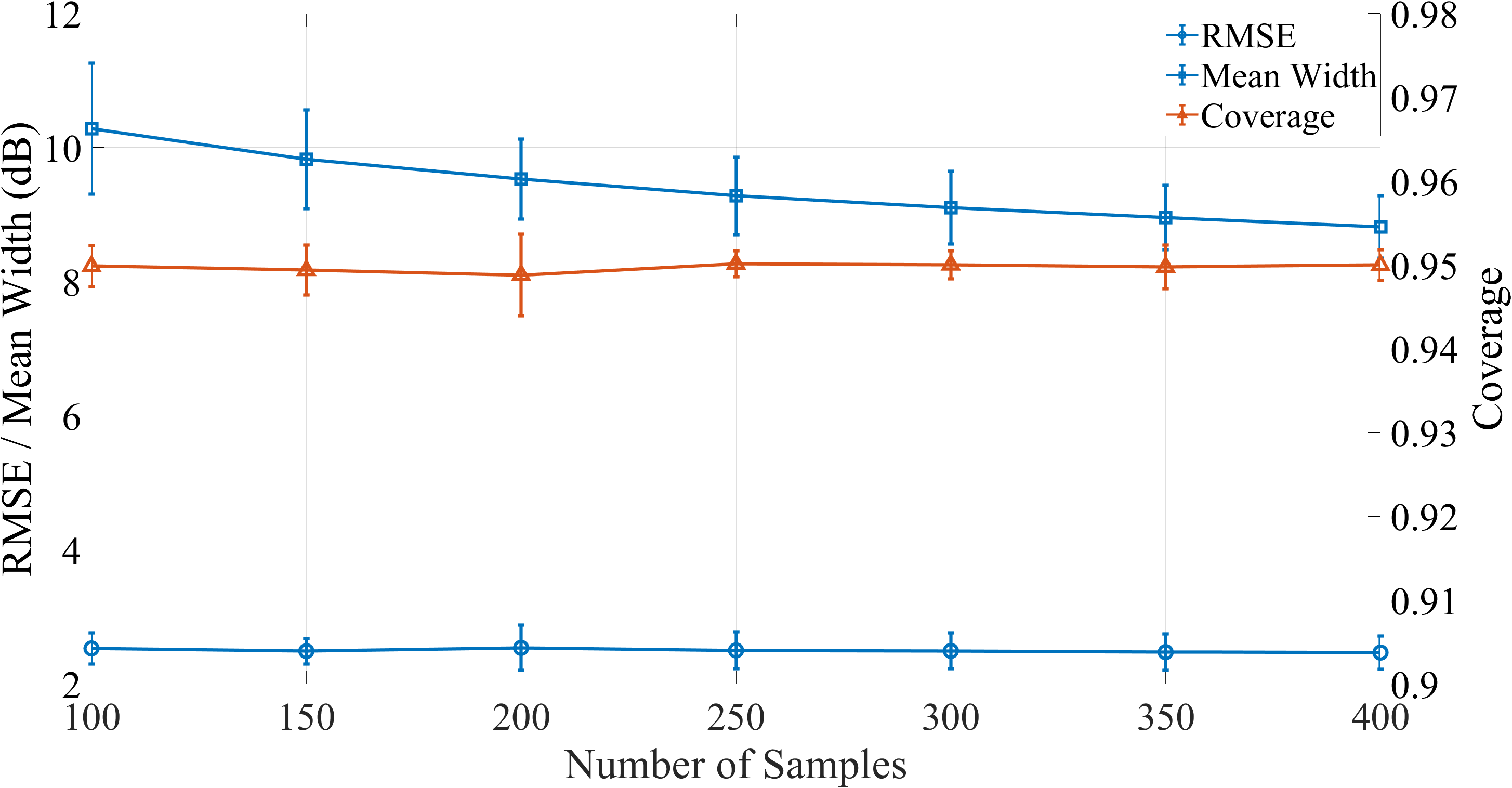}
\caption{Cross frequency residual correction using 4 and 5 GHz measurements for prediction at 4.5 GHz.}
\label{fig:cross_frequency_correction}
\end{figure}
 
The previous subsections analyze the WEKP constructed from RT simulation data. However, a mismatch may still exist between RT simulation and real measurement data. Therefore, a small number of measurement samples are used to correct the residual between the WEKP median prediction and the measured path loss. 
This subsection evaluates the residual correction method by examining the effect of the number of correction samples, comparing GPR with quantile regression (QR) under the same settings, and investigating cross-frequency prediction performance.

First, the influence of the number of correction samples is investigated to determine the amount of measurement data required for effective residual correction. 
For the correction sample selection, measurement samples from all frequency points are combined and randomly selected to construct the correction dataset. 
The selection process is repeated ten times, with the final results obtained by averaging the repeated experiments to reduce the influence of random sampling variations.
Figure~\ref{fig:number} presents the influence of correction sample size and the comparison between GPR and QR. 
As the number of correction samples increases, the RMSE of GPR gradually decreases, while the coverage consistently remains above the target coverage level. 
Meanwhile, the mean interval width remains relatively stable as the number of correction samples increases, indicating that the improved prediction performance is mainly attributed to more accurate residual estimation rather than wider prediction intervals.
Compared with QR, GPR achieves lower prediction errors and narrower prediction intervals under different correction sample sizes. 
For the selected correction sample size of 300, GPR provides a better balance between prediction accuracy and interval reliability than QR. 
The small variations among repeated random sampling experiments further indicate the stability of the correction process.
Although QR can directly estimate conditional quantiles, it requires wider intervals to maintain a comparable coverage level.

With the number of corrected samples fixed at 300, the correction performance is further analyzed at different frequency points.
The WEKP-PLP achieves the RMSE values of 1.17 dB, 1.29 dB, and 1.89 dB at 4 GHz, 4.5 GHz, and 5 GHz, respectively. 
Meanwhile, the coverage remains above or close to the target coverage level. 
The interval widths remain relatively stable among different frequencies, suggesting that the uncertainty characterization does not rely on excessively widening the prediction intervals. 
Therefore, the observed differences among frequencies mainly reflect the propagation characteristics rather than the instability of the correction method.

Figure~\ref{fig:interval} further visualizes the prediction intervals for randomly selected test samples at different frequencies. 
The blue shaded regions represent the prediction intervals, while the red dots indicate the measured path loss values. 
For 4 GHz and 4.5 GHz, most measurement samples are enclosed by the prediction intervals. 
At 5 GHz, some samples, especially those in the long distance region, fall outside the intervals. 
This is mainly related to the larger path loss fluctuation observed in the long distance measurements at 5 GHz, which increases the difficulty of residual correction. 
Nevertheless, the intervals still characterize the uncertainty variation of different propagation conditions and provide useful uncertainty information beyond the median path loss prediction.

The cross frequency correction experiment further evaluates the frequency generalization capability of the proposed framework. 
The measurement samples at 4 GHz and 5 GHz are randomly selected as the training samples, while the measurement samples at 4.5 GHz are used for performance evaluation.
As shown in Fig.~\ref{fig:cross_frequency_correction}, increasing the number of correction samples gradually improves the prediction accuracy, while the coverage consistently meets the target coverage requirement. 
Compared with the previous random sampling correction case, the intervals become wider under cross frequency correction. 
This increase in interval width is mainly caused by the additional uncertainty introduced by frequency transfer, since the residual at the source frequencies may not completely match that at the target frequency. 
However, the maintained coverage suggests that the proposed framework can effectively characterize the uncertainty and provide reliable prediction intervals when direct measurements at the target frequency are unavailable.

Overall, the measurement-based residual correction results indicate that a small number of measurement samples can effectively adapt the WEKP prior to the real propagation environment. The GPR method provides reliable point prediction and uncertainty characterization across different frequencies, while the cross frequency experiment further demonstrates its potential for the scenarios where only limited measurements are available at the target frequency.

\subsection{Comparison With Conventional Path Loss Models}

\begin{figure*}[!t]
    \centering
    \begin{subfigure}{0.31\linewidth}
        \centering
        \includegraphics[width=\linewidth]{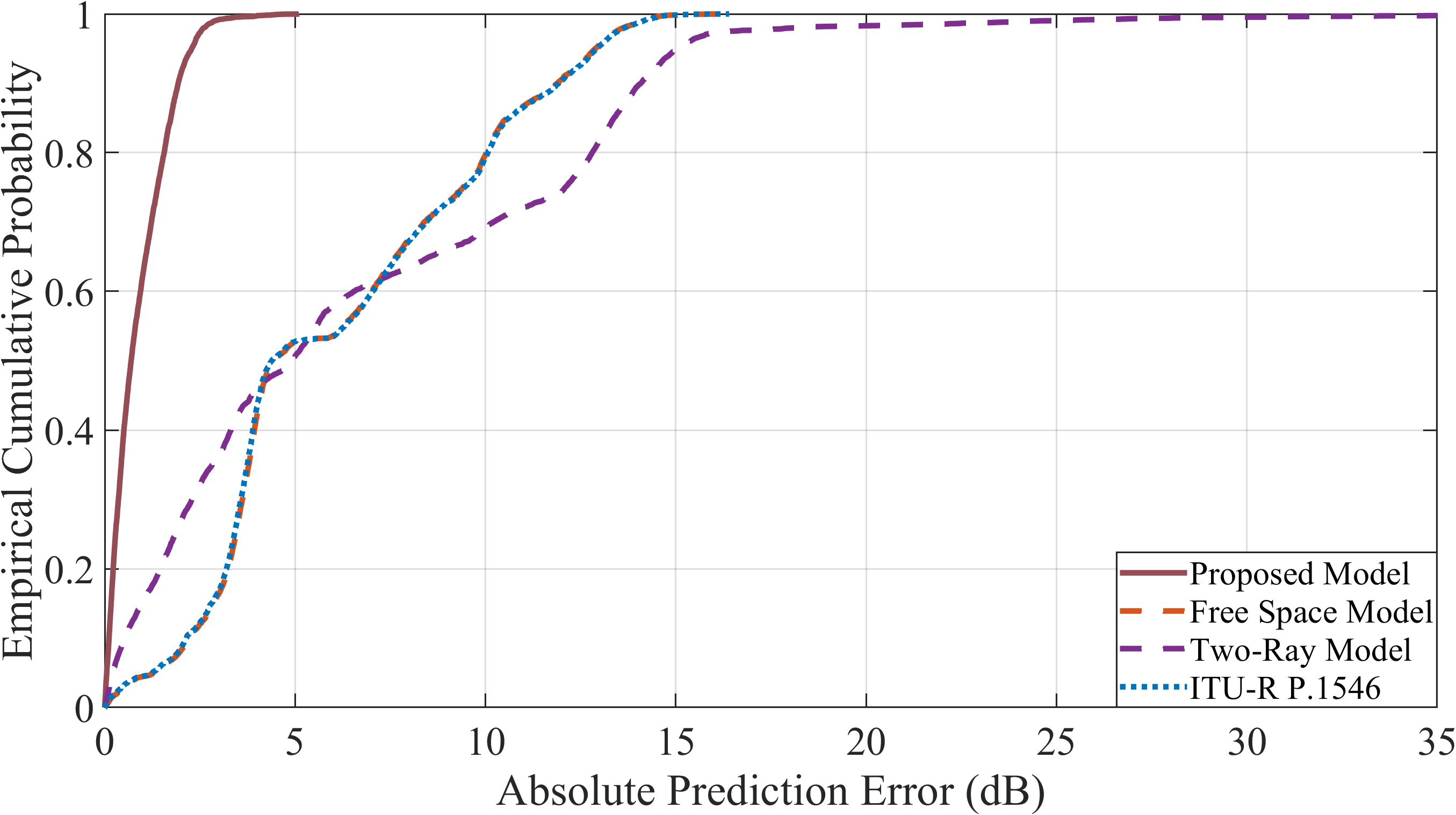} 
        \caption{}
    \end{subfigure}
    \hfill
    \begin{subfigure}{0.31\linewidth}
        \centering
        \includegraphics[width=\linewidth]{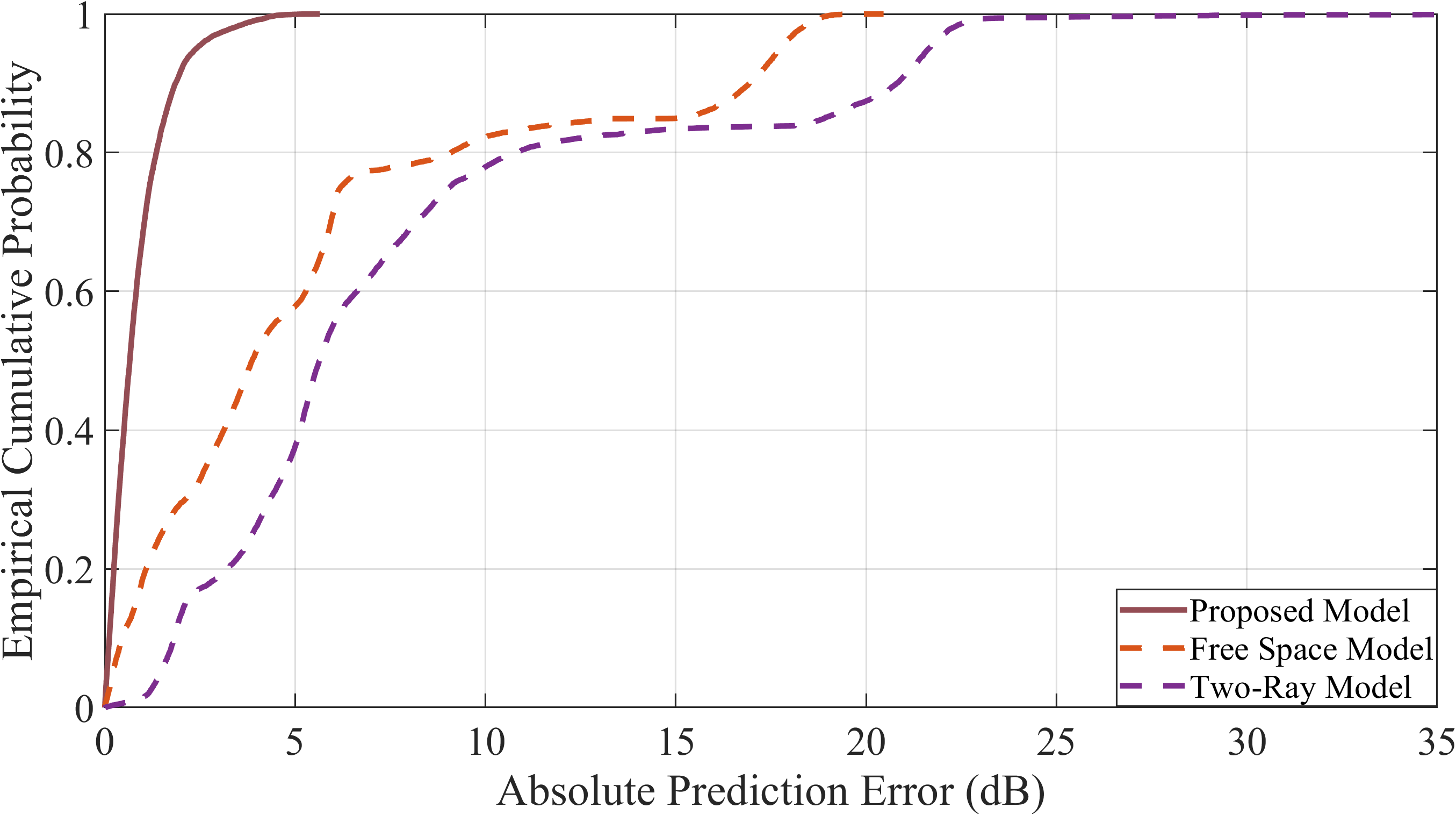} 
        \caption{}
    \end{subfigure}
    \hfill
    \begin{subfigure}{0.31\linewidth}
        \centering
        \includegraphics[width=\linewidth]{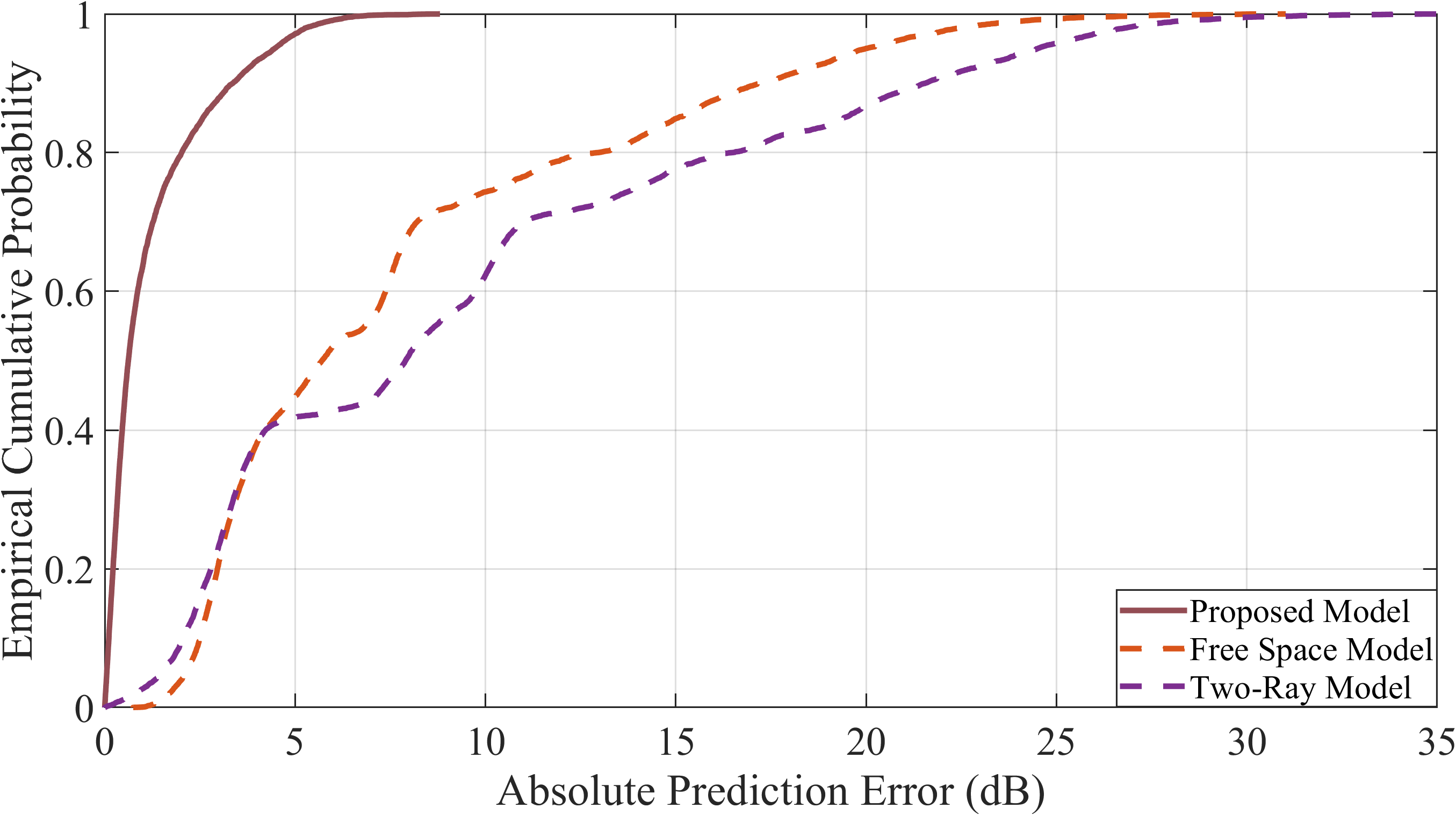} 
        \caption{} 
    \end{subfigure}
    \caption{ECDFs of the absolute prediction errors at different frequencies. (a) 4 GHz. (b) 4.5 GHz. (c) 5 GHz.}
    \label{fig10}
\end{figure*}

\begin{table}[!t]
    \centering 
    \caption{Comparison with conventional path loss models on the measurement dataset} 
    \label{tab:table6} 
    \renewcommand{\arraystretch}{1.1}
    \begin{tabular}{|c|c|c|c|} 
        \hline
        \makecell{\textbf{Model}} 
        & \makecell{\textbf{Evaluation Scope}} 
        & \makecell{\textbf{MAE}\textbf{(dB)}} 
        & \makecell{\textbf{RMSE}\textbf{(dB)}} \\
        \hline
        Proposed Method & All frequencies & 1.03 & 1.54 \\
        \hline
        FSPL  & All frequencies & 6.55 & 8.40 \\
        \hline
        Two-Ray  & All frequencies & 8.11 & 10.58 \\
        \hline
        ITU-R P.1546 & 4 GHz & 6.20 & 7.23 \\
        \hline
    \end{tabular}
\end{table}

To further evaluate the effectiveness of the proposed WEKP-PLP framework, its prediction performance is compared with that of conventional path loss models.
The compared models include the FSPL model, the two-ray model, and the ITU-R P.1546-6 model \cite{ITU_R_P1546_6}. 
The FSPL model considers only the free-space spreading loss determined by propagation distance and carrier frequency. 
The two-ray model considers the direct path and the sea surface reflected path under ideal propagation conditions. 
The ITU-R P.1546-6 model is an empirical propagation model that provides large-scale path loss prediction for terrestrial and coastal radio propagation scenarios \cite{zenodo}.
Since the applicable frequency range of ITU-R P.1546-6 is up to 4000 MHz, this model is evaluated only on the 4 GHz measurement subset. 
Therefore, ITU-R P.1546 is included only in the comparison at 4 GHz, whereas the proposed model, FSPL model, and two-ray model are evaluated at all three frequencies.

Table~\ref{tab:table6} compares the prediction errors of the proposed method and conventional path loss models. 
The proposed method achieves the lowest MAE and RMSE among the compared models. 
The large errors of FSPL model show that free-space attenuation alone can not accurately describe the measured shore-to-ship propagation.
Although the two-ray model considers the sea surface reflected path, its error remains large because the ideal deterministic two-path assumption can not fully capture the practical propagation variation in the measurement scenario. 
ITU-R P.1546-6 model provides an empirical reference, but its prediction error is still higher than that of the proposed method.
Figure~\ref{fig10} further presents the empirical cumulative distribution functions (ECDFs) of the absolute prediction errors at different frequencies. 
In an ECDF plot, a curve closer to the upper-left region means that more samples have smaller prediction errors. 
The proposed method shows a faster increase of cumulative probability than the conventional path loss models. This indicates that the proposed method reduces the prediction error for a larger portion of measurement samples, rather than improving only a few individual cases.

\subsection{Ablation Study of WEKP-PLP Framework}

\begin{table*}[t]
\centering
\caption{Ablation study of the proposed WEKP-PLP}
\label{tab:ablation}
\begin{tabular}{|c|c|c|c|c|c|c|}
\hline
\textbf{Method} & \textbf{MAE (dB)} & \textbf{RMSE (dB)} & $\bm{R^2}$  & \textbf{Coverage} & \textbf{Mean Width (dB)} & \textbf{Max Width (dB)} \\
\hline
WEKP Prior Only & 5.44 & 7.27 & 0.43 & 51.26\% & 10.04 & 11.55 \\
\hline
Two-Ray Prior and GPR Correction & 2.95 & 5.12  & 0.79 & 92.40\% & 16.22 & 22.10 \\
\hline
WEKP Prior and GPR Correction & 1.03 & 1.54 & 0.98 & 94.09\% & 6.06 & 9.41 \\
\hline
\end{tabular}
\end{table*}

The ablation study analyzes the roles of the WEKP prior and measurement-based residual correction in this subsection. 
Specifically, it evaluates whether the WEKP prior can be directly applied to the measurement scenario without correction, and whether the WEKP provides a more informative prior than the two-ray model when the same GPR based residual correction is applied.

As shown in Table~\ref{tab:ablation}, directly applying the WEKP prior to the measurement scenario leads to large prediction errors and low coverage. This suggests that the WEKP prior can not be directly used as the final prediction in the measurement scenario. The mismatch is expected because the RT simulation mainly considers the direct path and the sea surface reflected components, while real measurements may contain additional propagation effects.
When the same residual correction is applied, the WEKP prior leads to much better performance than the two-ray prior. The two-ray prior with GPR still has a large RMSE and a wide interval, while the WEKP prior with GPR obtains a much lower RMSE and a more compact interval. This suggests that the quality of the prior affects the performance of residual correction. A weak prior is difficult to correct using a small number of measurement samples.

Overall, the ablation results indicate that both components are important for the proposed framework. The WEKP prior provides point prediction and interval information, while the residual correction adapts this prior to real shore-to-ship scenarios. 

\section{Conclusion} \label{sec:6}

This paper proposed a WEKP-PLP framework for shore- to-ship path loss prediction with uncertainty characterization. 
The proposed framework first constructs a WEKP prior from RT simulation data, and then uses measurement samples to correct the residual between the WEKP median prediction and the measured path loss. 
Different from conventional path loss models, the proposed framework provides both median path loss prediction and corresponding uncertainty information.
The experimental results demonstrate that the WEKP can capture the residual behavior between RT simulation and the FSPL baseline, and provide adaptive uncertainty information under different propagation conditions. 
After measurement-based residual correction, the proposed method achieves accurate path loss prediction and reliable uncertainty characterization on the measurement dataset. 
With a correction sample size of 300, the proposed framework achieves an MAE of 1.03 dB and an RMSE of 1.54 dB, while achieving the target coverage level.
The ablation results confirm that both the WEKP prior and the measurement-based residual correction are necessary. 
The future work will incorporate more diverse sea states and independent measurement routes to further validate the prediction capability of the proposed framework.


\begin{thebibliography}{00}

\bibitem{jianhua_2023} J. Zhang et al., ``6G channel modeling: Requirement, measurement, methodology and simulator,'' \textit{arXiv preprint arXiv:2305.16616}, May 2023.

\bibitem{Bose_2025} P. Bose, G. Rafiq, and P. Orten, ``Maritime communications--A review of potential wireless communication technologies,'' \textit{Ocean Eng.}, vol. 341, Art. no. 122527, 2025.

\bibitem{zhang_2024} J. Zhang, H. Wang, Y. Zhang, et al., ``Channel characteristics and modeling research for 6G: challenges, progress, and prospects,'' \textit{Sci. China Inf. Sci.}, vol. 54, no. 5, pp. 1114--1143, 2024.

\bibitem{Xylouris_2024} G. Xylouris, N. Nomikos, A. Kalafatelis, A. Giannopoulos, S. Spantideas, and P. Trakadas, ``Sailing into the future: Technologies, challenges, and opportunities for maritime communication networks in the 6G era,'' \textit{Front. Commun. Netw.}, vol. 5, Art. no. 1439529, Jul. 2024.

\bibitem{wang_2018} J. Wang, H. Zhou, Y. Li, et al., ``Wireless channel models for maritime communications,'' \textit{IEEE Access}, vol. 6, pp. 68070--68088, 2018.

\bibitem{tian_2026} L. Tian, J. Du, Z. Ding, J. Zhang, Y. He, and L. Guo, ``BUPTCMCC-6G-DataAI-Maritime: A configurable channel dataset of offshore scenario for AI research,'' \textit{Sci. China Inf. Sci.}, vol. 69, no. 4, 2026.

\bibitem{wangh_2025} H. Wang, J. Zhang, G. Nie, et al., ``Digital twin channel for 6G: Concepts, architectures and potential applications,'' \textit{IEEE Commun. Mag.}, vol. 63, no. 3, pp. 24--30, Mar. 2025.

\bibitem{mi_2025} Y. Mi, X. Zhang, L. Kong, et al., ``Measurement-based characterization and modeling of land-to-ship wireless channels at 3.2 GHz,'' \textit{IEEE Trans. Antennas Propag.}, vol. 73, no. 1, pp. 441--452, Jan. 2025.

\bibitem{sun_2026} S. Sun, Y. Guo, M. Tao, et al., ``Modeling and analysis of land-to-ship maritime wireless channels at 5.8 GHz,'' \textit{IEEE Trans. Wireless Commun.}, vol. 25, pp. 10051--10065, 2026.

\bibitem{yang_2019} K. Yang, A. F. Molisch, T. Ekman, T. R{\o}ste, and M. Berbineau, ``A round earth loss model and small-scale channel properties for open-sea radio propagation,'' \textit{IEEE Trans. Veh. Technol.}, vol. 68, no. 9, pp. 8449--8460, Sep. 2019.

\bibitem{wang_2025} Y. Wang, T. Zhou, W. Feng, T. Xu, and H. Hu, ``When maritime wireless communications meet evaporation ducts: A three-ray path loss modeling perspective,'' \textit{IEEE Wireless Commun. Lett.}, vol. 14, no. 2, pp. 470--474, Feb. 2025.

\bibitem{jiaqi_2025} J. Zhang, L. Tian, P. Tang, et al., ``Measurement and analysis of wideband wireless channel in shore-to-ship scenarios,'' in \textit{Proc. IEEE Veh. Technol. Conf. (VTC-Fall)}, Chengdu, China, 2025, pp. 1--6.

\bibitem{ding_2019} R. Ding, J. Wang, Y. Li, L. You, and Q. Sun, ``A stochastic ray-tracing approach for maritime line-of-sight channel modeling,'' in \textit{Proc. IEEE Int. Conf. Commun. Technol. (ICCT)}, Xi'an, China, 2019, pp. 135--139.

\bibitem{chen_2022} Z. Chen, L. Wang, and R. Chen, ``Offshore electromagnetic wave propagation loss model based on ray tracing method,'' in \textit{Proc. Photon. Electromagn. Res. Symp. (PIERS)}, Hangzhou, China, 2022, pp. 171--176.

\bibitem{wang_2021} L. Wang, Z. Chen, and Y. Zhang, ``Acceleration of offshore electromagnetic energy distribution prediction algorithm based on ray-tracing method and PM wave spectrum,'' \textit{IEEE Antennas Wirel. Propag. Lett.}, vol. 20, no. 12, pp. 2363--2367, Dec. 2021.

\bibitem{yu_2025} L. Yu, L. Shi, J. Zhang, Z. Zhang, Y. Zhang, and G. Liu, ``ChannelGPT: A large model toward real-world channel foundation model for 6G environment intelligence communication,'' \textit{IEEE Commun. Mag.}, vol. 63, no. 10, pp. 68--74, Oct. 2025.

\bibitem{noh_2024} E. Noh and J. Park, ``A two-stage model for enhanced prediction of received signal strength in maritime environments,'' \textit{IEEE Antennas Wirel. Propag. Lett.}, vol. 23, no. 8, pp. 2531--2535, Aug. 2024.

\bibitem{zhangh_2024} H. Zhang, T. Zhou, T. Xu, M. Cheng, and H. Hu, ``Field measurement and channel modeling around Wailingding Island for maritime wireless communication,'' \textit{IEEE Antennas Wireless Propag. Lett.}, vol. 23, no. 6, pp. 1934--1938, Jun. 2024.

\bibitem{peng_2025} L. S. Peng, K. Yang, J. M. Wu, et al., ``Machine learning methods comparison for maritime wireless signal strength prediction,'' \textit{Eng. Appl. Artif. Intell.}, vol. 158, Art. no. 111357, Oct. 2025.

\bibitem{jialin_2024} J. Wang, J. Zhang, Y. Sun, et al., ``Electromagnetic wave property inspired radio environment knowledge construction and artificial intelligence based verification for 6G digital twin channel,'' \textit{Front. Inf. Technol. Electr. Eng.}, vol. 26, no. 2, pp. 260-278, Feb. 2025.

\bibitem{jialin_2025} J. Wang, J. Zhang, Y. Zhang, et al., ``Radio environment knowledge pool for 6G digital twin channel,'' \textit{IEEE Commun. Mag.}, vol. 63, no. 5, pp. 158--164, May. 2025.

\bibitem{jiaxin_2024} J. Zhang, J. Lin, P. Tang, et al., ``Deterministic ray tracing: A promising approach to THz channel modeling in 6G deployment scenarios,'' \textit{IEEE Commun. Mag.}, vol. 62, no. 2, pp. 48--54, Feb. 2024.

\bibitem{toporkov_2000} J. V. Toporkov and G. S. Brown, ``Numerical simulations of scattering from time-varying, randomly rough surfaces,'' \textit{IEEE Trans. Geosci. Remote Sens.}, vol. 38, no. 4, pp. 1616--1625, Jul. 2000.

\bibitem{jiang_2015} W.-Q. Jiang, M. Zhang, P.-B. Wei, and D. Nie, ``Spectral decomposition modeling method and its application to EM scattering calculation of large rough surface with SSA method,'' \textit{IEEE J. Sel. Top. Appl. Earth Obs. Remote Sens.}, vol. 8, no. 4, pp. 1848--1854, Apr. 2015.

\bibitem{pierson_1964} W. J. Pierson, Jr. and L. Moskowitz, ``A proposed spectral form for fully developed wind seas based on the similarity theory of S. A. Kitaigorodskii,'' \textit{J. Geophys. Res.}, vol. 69, no. 24, pp. 5181--5190, Dec. 1964.

\bibitem{Meissner_2004} T. Meissner and F. J. Wentz, ``The complex dielectric constant of pure and seawater from microwave satellite observations,'' \textit{IEEE Trans. Geosci. Remote Sens.}, vol. 42, no. 9, pp. 1836--1849, Sep. 2004.

\bibitem{huang_2016} F. Huang, X. Liao, and Y. Bai, ``Multipath channel model for radio propagation over sea surface,'' \textit{Wireless Pers. Commun.}, vol. 90, no. 1, pp. 245--257, Sep. 2016.

\bibitem{Prokhorenkova_2018} L. Prokhorenkova, G. Gusev, A. Vorobev, et al., ``CatBoost: Unbiased boosting with categorical features,'' in \textit{Proc. Adv. Neural Inf. Process. Syst. (NeurIPS)}, vol. 31, 2018.

\bibitem{miao_2023} H. Miao, J. Zhang, P. Tang, et al., ``Sub-6 GHz to mmWave for 5G-Advanced and beyond: Channel measurements, characteristics and impact on system performance,'' \textit{IEEE J. Sel. Areas Commun.}, vol. 41, no. 6, pp. 1945--1960, Jun. 2023.

\bibitem{Williams_1995}
C. K. I. Williams and C. E. Rasmussen, ``Gaussian processes for regression,'' in \textit{Adv. Neural Inf. Process. Syst. (NeurIPS)}, vol. 8, 1995.

\bibitem{Jang_2022}
K. J. Jang et al., ``Path Loss Model Based on Machine Learning Using Multi-Dimensional Gaussian Process Regression," \textit{IEEE Access}, vol. 10, pp. 115061-115073, 2022.

\bibitem{Lundberg_2017} S. M. Lundberg and S.-I. Lee, ``A unified approach to interpreting model predictions,'' in \textit{Proc. Adv. Neural Inf. Process. Syst. (NeurIPS)}, vol. 30, 2017.

\bibitem{sun_2022} Y. Sun, J. Zhang, Y. Zhang, et al., ``Environment features-based model for path loss prediction,'' \textit{IEEE Wireless Commun. Lett.}, vol. 11, no. 9, pp. 2010--2014, Sep. 2022.

\bibitem{ITU_R_P1546_6} ITU-R, ``Method for point-to-area predictions for terrestrial services in the frequency range 30 MHz to 4 000 MHz,'' \textit{Recommendation ITU-R P.1546-6}, International Telecommunication Union, Geneva, Switzerland, Aug. 2019.

\bibitem{zenodo} Ivica Stevanovic, ``MATLAB/Octave Implementation of Recommendation ITU-R P.1546,'' \textit{Zenodo}, Feb. 18, 2022.
\end{thebibliography}
\end{document}